\documentclass[journal=apchd5,manuscript=article,layout=twocolumn]{achemso}
\usepackage{array,booktabs}
\usepackage{indentfirst}
\usepackage{amsmath}
\newcolumntype{P}[1]{>{\centering\arraybackslash}p{#1}}

\usepackage[version=3]{mhchem} % Formula subscripts using \ce{}

\author{Kristen M. Parzuchowski}
\email{kristen.parzuchowski@nist.gov}
\affiliation[JILA]
{JILA, University of Colorado Boulder, Colorado 80309, USA}
\alsoaffiliation[PhysDep]
{Department of Physics, University of Colorado Boulder, Colorado 80309, USA}
\alsoaffiliation[aNIST]
{Associate of the National Institute of Standards and Technology, Boulder, Colorado 80305, USA}
\author{Michael D. Mazurek}
\affiliation[aNIST]
{Associate of the National Institute of Standards and Technology, Boulder, Colorado 80305, USA}
\alsoaffiliation[PhysDep]
{Department of Physics, University of Colorado Boulder, Colorado 80309, USA}
\author{Charles H. Camp Jr.}
\affiliation[NIST G]
{National Institute of Standards and Technology, Gaithersburg, Maryland 20899, USA}
\author{Martin J. Stevens}
\affiliation[NIST]
{National Institute of Standards and Technology, Boulder, Colorado 80305, USA}
\author{Ralph Jimenez}
\email{rjimenez@jila.colorado.edu}
\affiliation[JILA]
{JILA, University of Colorado Boulder, Colorado 80309, USA}
\alsoaffiliation[ChemDep]
{Department of Chemistry, University of Colorado Boulder, Colorado 80309, USA}
\title[An \textsf{achemso} demo]
    {A Liquid-Core Fiber Platform for Classical and Entangled Two-Photon Absorption Measurements}


\begin{document}
%%%%%%%%%%%%%%%%%%%%%%%%%%%%%%%%%%%%%%%%%%%%%%%%%%%%%%%%%%%%%%%%%%%%%
\begin{tocentry}
\begin{center}
\includegraphics[width=8.25cm,height=4.45cm]{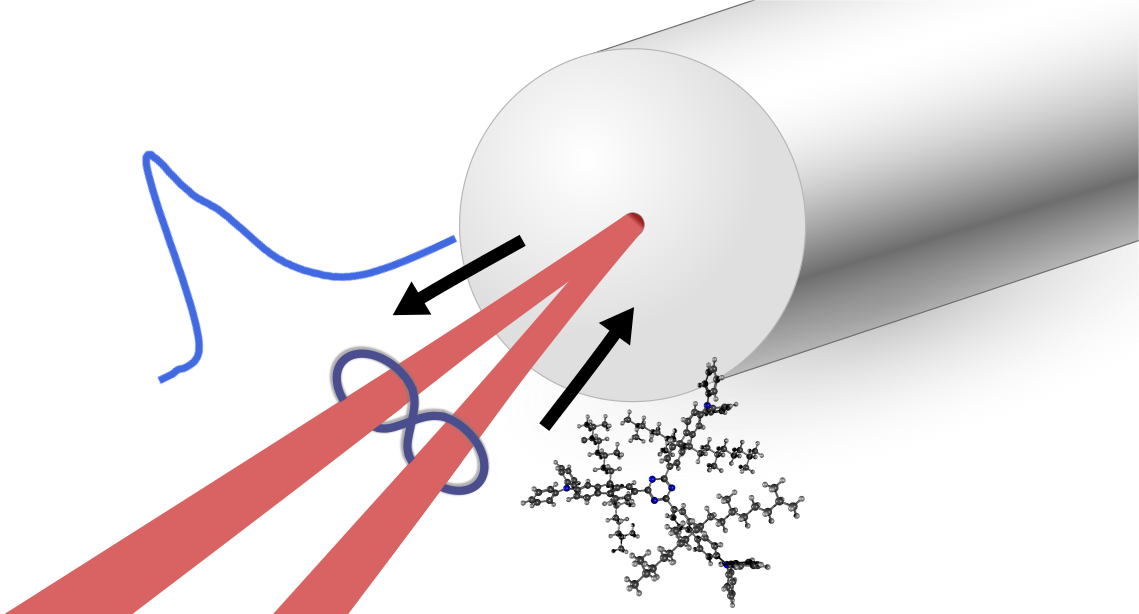}
%this text explains the graphic
\end{center}
%\newline
Correlated photon pairs (red) are coupled into a fiber containing a solution of the molecule AF455. Fluorescence (blue) generated from the two-photon absorption process is guided out of the fiber. 
\end{tocentry}
%%%%%%%%%%%%%%%%%%%%%%%%%%%%%%%%%%%%%%%%%%%%%%%%%%%%%%%%%%%%%%%%%%%%%
%%%%%%%%%%%%%%%%%%%%%%%%%%%%%%%%%%%%%%%%%%%%%%%%%%%%%%%%%%%%%%%%%%%%%
\begin{abstract}
We introduce a toluene-filled fiber platform for two-photon absorption measurements. By confining both the light and molecular sample inside the 5~$\mu$m hollow core of the fiber, we increase the distance over which the nonlinear light-matter interaction occurs. With only a 7.3~nL excitation volume, we measure classical two-photon absorption (C2PA) at an average laser power as low as 1.75~nW, which is a 45-fold improvement over a conventional free-space technique. We use this platform to attempt to measure entangled two-photon absorption (E2PA), a process with a limited regime where the quantum advantage is large. This regime arises due to a crossover from linear to quadratic scaling with photon flux as photon flux is increased. Recently, several teams of researchers have reported that E2PA cross-sections are much smaller than previously claimed. As a result, the linear scaling dominates at photon fluxes so low that it is extremely difficult or impossible to measure using conventional free-space techniques. In this report, we implement the first E2PA measurement using a waveguide. We see no evidence of E2PA, and we set an upper bound on the cross-section consistent with these recent reports. 
\end{abstract}

%%%%%%%%%%%%%%%%%%%%%%%%%%%%%%%%%%%%%%%%%%%%%%%%%%%%%%%%%%%%%%%%%%%%%

\section*{Introduction}
Liquid-core fibers (LCF) provide platforms for material characterization\cite{Dallas2004}, optical limiting\cite{Khoo2004}, sensing\cite{Herraez2015} and light generation\cite{Gerosa2015,Scheibinger2021}. These fibers are designed using a variety of geometries and materials that aid in their ability to confine light into micrometer-scale cores over centimeter-scale distances. The light reaches high intensities within LCF, which enables nonlinear processes to dominate\cite{Bhagwat2008,Chemnitz2023} and thus become easier to observe than in conventional free-space techniques. Processes like four-wave mixing, third-harmonic generation, supercontinuum generation, Raman scattering and two-photon absorption (2PA) have been targeted in LCF. 
 
Two-photon absorption was first observed in capillary LCFs in the 1990s\cite{He1995,Khoo1996}. These studies included models that were used to derive molecular 2PA coefficients from measurements of energy transmitted out of the fiber as a function of energy sent into the fiber\cite{He1997,Khoo1998}. A number of 2PA fiber studies have followed these, including 2PA observed in liquid- or vapor-core photonic crystal fibers\cite{Williams2014,Saha2011} and using tapered or exposed-core fiber\cite{Hendrickson2010,Perrella2015}. One study in particular\cite{Williams2014} used a water-filled photonic-crystal-fiber system to measure a 2PA signal at analyte concentrations as low as $10^{-9}$~M (M = mol~L$^{-1}$), which is roughly four orders-of-magnitude smaller than typical concentrations used in free-space configurations. Although a one-to-one comparison of the two techniques was not performed, this report exemplifies how sensitive LCF techniques can lead to significant improvements in 2PA signal levels.

Here, we show that with a toluene-filled fiber, 2PA cross-sections, which are the quantities that describe the strength of the 2PA process, can be measured at extremely low powers. To the best of our knowledge, this work is the first to show a 2PA measurement at an average power as low as 1.75~nW. This is a 45-fold lower power than that achieved in our previous results\cite{Parzuchowski} for low-power 2PA in free space for the same sample. These fiber and free-space measurements serve as a one-to-one comparison since both were designed to test the sensitivity limits of the platforms. We believe this is the first time that the advantage of the fiber platform for 2PA has been quantified. 

%Furthermore, the platform can be used for detecting low analyte concentrations and requires a sample volume of only 7.3~nL.
This sensitivity, paired with the ability to increase our laser power by up to nine orders of magnitude, demonstrates that the platform is well equipped to boost previously undetectable signals. This is necessary, for example, for measurements of C2PA cross-sections at much less efficient wavelengths or for much less efficient absorbers. Furthermore, rather than being limited to a high peak power laser\cite{Rumi2010}, unconventional light sources may produce sizeable 2PA signals.

One such light source is an entangled photon pair source based on spontaneous parametric downconversion (SPDC). We employ an SPDC source in an attempt to measure entangled two-photon absorption (E2PA). To differentiate between E2PA and 2PA excited by a laser source, we refer to the latter process as classical two-photon absorption (C2PA). The E2PA process has been a subject of theoretical study since the 1980s (Refs.~\citenum{Javanainen1990,Gea-Banacloche1989,Fei1997}), driven by the idea that the strong temporal and spatial correlations of SPDC photons are ideal for 2PA. These temporal and spatial correlations are characterized by the entanglement time and area, respectively\cite{Fei1997}. The entanglement time quantifies the width of the joint probability distribution of differences in arrival times of the two photons forming a pair, defined at the location of interest. The entanglement area is proportional to the width of the joint probability distribution of differences in positions of the two photons forming a pair, defined at the location of interest---to calculate entanglement area, the widths from the two transverse dimensions must be used. 

The E2PA process is known to scale linearly with the excitation photon flux, rather than the quadratic scaling of C2PA, in the low-gain parametric downconversion (PDC) regime where, if it can be measured, it dominates over the quadratic component and thus demonstrates a quantum advantage. In the high-gain PDC regime, a quadratic scaling is recovered, with the possibility of up to three-fold signal boost relative to excitation with a coherent laser source. This boost reflects a higher likelihood of two photons overlapping due to the increased statistical fluctuations of the excitation source. A more extensive description of the SPDC regimes and expected quantum advantage for various bandwidth conditions of the pump, SPDC and absorption spectrum is covered in Refs.~\citenum{Raymer2022,Landes2024}.

Recently, a wide range of evidence from a variety of experimental research studies has shown that molecular E2PA cross-sections are much smaller than previously reported. The majority of studies have reported no E2PA signal above the noise floor in the low-gain (Refs.~\citenum{Mikhaylov2020,Mazurek,Landes2021,Mikhaylov2022,CoronaAquino2022,Hickam2022,Arango2023,Gabler2023,Qian2024,He2024,Landes2024}) and moderate-gain PDC regimes (Refs.~\citenum{Parzuchowski}). One study reported an extremely weak linear-scaling signal (Ref.~\citenum{Tabakaev2021}), however another study closely replicated this experiment and was unable to observe the reported signal\cite{Landes2024}. Two studies have reported quadratic-scaling signals using PDC (Ref.~\citenum{Qian2024,Landes2024}). In the report by Raymer and coworkers~\cite{Landes2024}, a high-gain PDC source is used and the resulting signal exhibits a moderate boost relative to a coherent source, which is consistent with the increased photon number fluctuations of the excitation source. In the report by Wang and coworkers~\cite{Qian2024}, a low-gain PDC source is used to excite upconversion nanoparticles that have long-lived intermediate states and thus can be excited via uncorrelated photon pairs despite using an excitation source with significant temporal separation between photon pairs. These experimental reports are reinforced by recent theoretical works that show that the linear scaling will dominate at photon fluxes so low that the resulting minuscule signal will be difficult or impossible to measure with current technology (Refs.~\citenum{Raymer2021, Landes2021_2, Raymer2022, Drago}). Furthermore, the conditions under which the minuscule signal is maximized require careful selection of the SPDC source and two-photon absorber. These experimental and theoretical studies all consider conventional free-space excitation conditions. 

Here we present, to the best of our knowledge, the first E2PA measurement implemented in a waveguide. Our approach uses a relatively high flux of photon pairs to maximize our likelihood of measuring a signal. In our setup, this high flux is generated by a moderate-gain PDC process. We do not observe E2PA and set an upper bound on the E2PA cross-section of the molecule AF455. 

In this paper, we discuss the principles of our experimental scheme and the design considerations along with our experimental implementation and its characterization. Next, we present our results for excitation of the molecule AF455 with both laser and SPDC photons and derive the C2PA and E2PA cross-sections, respectively. Finally we compare our results with those from previous free-space measurements\cite{Parzuchowski}. More details on experimental characterization, data acquisition and data analysis are provided in the supplemental information.

\section*{Operating Principles}
\label{Sec:FiberOpPrinc}

We confine both a sample (two-photon-absorbing molecules in solution) and an excitation beam in the core of a hollow-core fiber. Here the goal is to excite the confined molecules by 2PA. The excited molecules sometimes emit fluorescence and some of that light is guided back out of the fiber and can be detected.

\begin{figure*}[!tbh]
\centering
\includegraphics[width=0.99\textwidth,trim=15 5 5 5, clip]{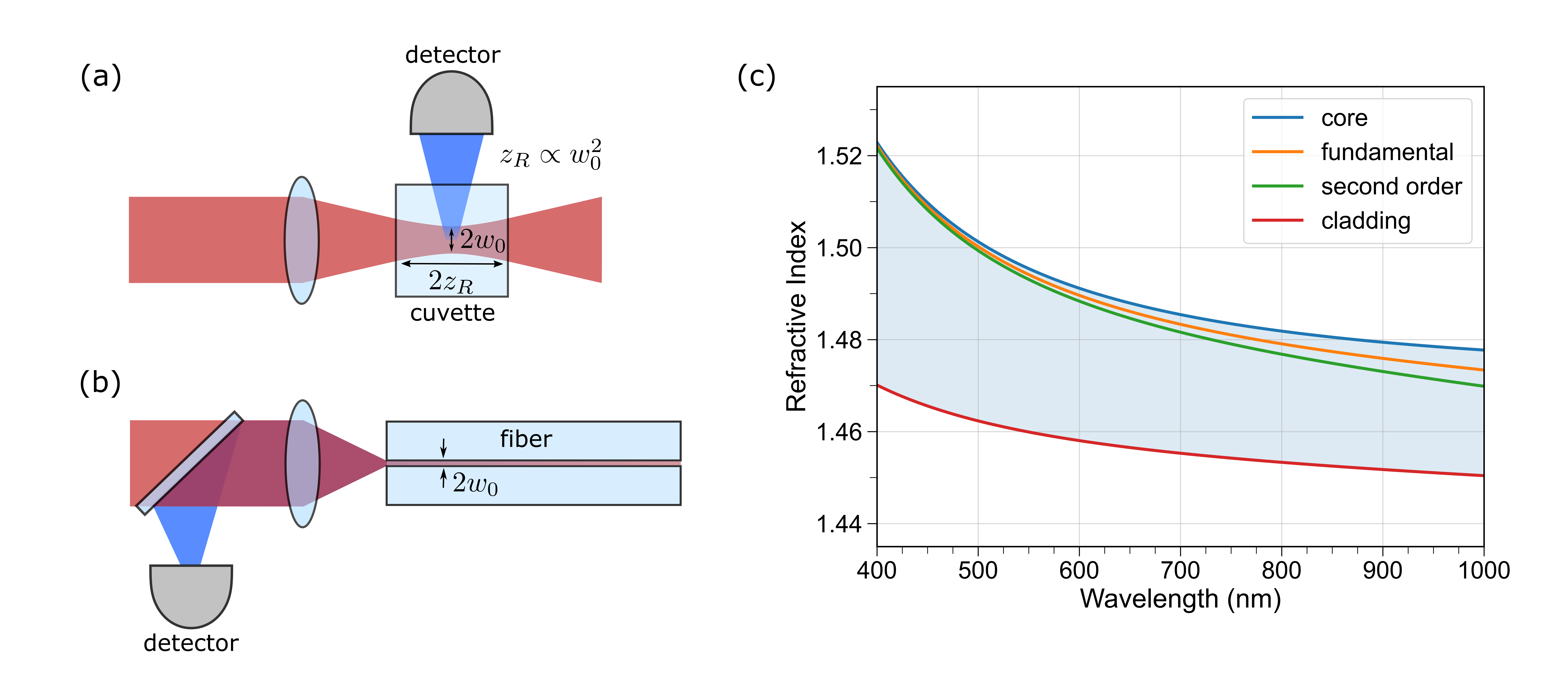}
\caption[Advantage of fiber-based measurement]{\label{fig:CuvetteVsFiberAndIndexDiagram}Illustration of (a) a conventional free-space and (b) a fiber-based 2PA measurement scheme. In the free-space approach, the light is focused into the cuvette to a beam waist $w_0$. The Rayleigh range $z_R$ characterizes the distance over which the beam size is $\le \sqrt{2}w_0$. When $w_0$ is decreased so is $z_R$, thus high intensity beams can only be maintained over short distances. Fluorescence is generated primarily at the waist of the beam and the fraction emitted in the direction of the detector is collected. In the fiber-based approach, the light is focused into fiber with waist $w_0$ and maintained over long lengths. Fluorescence is generated along the length of the fiber and the fraction directed opposite of the excitation beam is collected on a detector. (c) Refractive index as a function of wavelength for the relevant media and first two modes of guided light. The toluene-filled core (blue) and silica cladding (red) indices are shown. The mode indices for the fundamental (orange) and second-order (green) modes are shown. Additional modes can propagate through the filled fiber with mode indices filling the blue shaded region between the second-order mode and cladding indices.}
\end{figure*}

The benefit of this scheme is illustrated in Fig.~\ref{fig:CuvetteVsFiberAndIndexDiagram}(a-b). In the free-space approach (Fig.~\ref{fig:CuvetteVsFiberAndIndexDiagram}(a)), light is focused into the cuvette to a waist $w_0$ and diverges on a length scale characterized by the Rayleigh range $z_R$ which scales quadratically with $w_0$. Thus, increasing beam intensity occurs at the cost of maintaining the intensity over a shorter distance $z_R$. For typical focusing conditions of $w_0\approx 30\,\mu$m for an 810~nm beam into toluene, $z_R\approx5.2$~mm, which is about half of the cuvette's length. In contrast, with the fiber-based approach (Fig.~\ref{fig:CuvetteVsFiberAndIndexDiagram}(b)), the beam waist $w_0$ is maintained with low loss over the entire propagation length of the fiber because of optical confinement. Typically, $w_0$ is much smaller in fiber than in free space. This high light intensity is advantageous for 2PA because two photons must be spatially overlapped at a molecule for absorption to occur. Maintaining the peak intensity of the focused beam over centimeter length-scales increases the excitation probability because of the interaction with a large number of molecules.

 To achieve broadband guidance of both the excitation and fluorescence photons, we selected our fiber for index guidance. In this regime, nearly all light traveling in the core and incident onto the core-cladding interface at an angle (with respect to the normal to the interface) equal to or greater than the critical angle, $\theta_c = \textrm{arcsin}(n_\mathrm{clad}/n_\mathrm{core})$ where $n_\mathrm{clad}$ and $n_\mathrm{core}$ are the indices of refraction of the cladding and core of the fiber, is guided through the fiber by total internal reflection. It is likely that our cladding material is imperfect, due in part to optical losses and its finite extent, and results in a minor deviation from perfect total internal reflection. The critical angle is only a real number if the refractive index of the cladding is smaller than that of the core. To satisfy this criterion, we use a standard capillary tubing (inner diameter = $5\,\mu$m, outer diameter = $150\,\mu$m) with a silica cladding and fill the hollow core with toluene. The critical angle under these conditions is real for both the excitation and fluorescence wavelength regions. Many other common solvents, such as water, methanol or chloroform, have indices of refraction smaller than that of silica.

The refractive indices as a function of wavelength for the toluene core and silica cladding are plotted in Fig.~\ref{fig:CuvetteVsFiberAndIndexDiagram}(c) using the known dispersion equations for the materials\cite{Moutzouris2013,Malitson1965}. The effective index of refraction of the fundamental and second-order modes are shown and are calculated using Ref.~\citenum{Fini2004}. We can estimate the number of modes that can propagate along the fiber, which has a core diameter $d$, using the V-number,
\begin{equation}
\label{eq:Vnum}
    V(\lambda) = \frac{\pi d}{\lambda} \sqrt{n_\mathrm{core}^2(\lambda)-n_\mathrm{clad}^2(\lambda)}.
\end{equation}
The number of modes that can propagate at a particular wavelength $\lambda$ is then $V^2(\lambda)/2$. For the excitation wavelength of 810~nm, approximately 16 modes can propagate in our LCF. For the fluorescence wavelengths of AF455 in toluene, which is peaked at 451~nm, approximately 80 modes can propagate. In the ideal case, all the light would remain in the fundamental mode because this mode experiences the lowest loss, has lower dispersion and has a Gaussian spatial profile. All these characteristics will increase the rate of 2PA.  
\begin{figure*}[!t]
\includegraphics[width=0.99\textwidth]{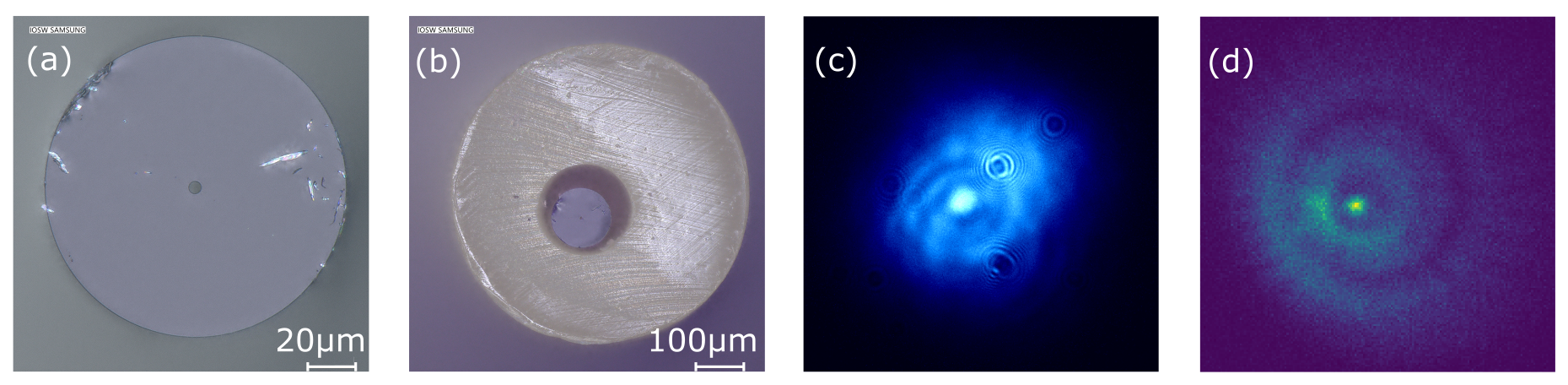}
\caption[Images of the fiber]{\label{fig:Capillary Images}Digital microscope images of one end of (a) the fiber and (b) the fiber inside of the tubing sleeve. (c) sCMOS image of the 810~nm excitation laser guided through the fiber. (d) EMCCD image of the fluorescence from AF455 guided out of the fiber. The last two images are taken in the image plane of the fiber face. Some features of the fiber and tubing sleeve are visible in both images as described in the Supporting Information.}
\end{figure*}

In addition to toluene allowing broadband guidance of light in a silica fiber, this solvent also has a low absorption coefficient\cite{Kedenburg2012} at the excitation and fluorescence wavelengths (0.0030~cm$^{-1}$ and 0.0039 cm$^{-1}$, respectively). This is a necessary condition otherwise the long length of the fiber will add little to no benefit since all the light would be absorbed after a short distance. For comparison, water has an absorption coefficient that is about one order of magnitude larger at 810~nm (0.0209~cm$^{-1}$\cite{Kedenburg2012}). 

\section*{Methods}
To prepare the fiber, we first cut clean facets on both ends to minimize light-coupling losses. We use a custom-built coil heater to remove about 20~mm of the polyamide coating from each end, then we cleave the fiber ends in the regions where the coatings were removed. We inspect the ends under a digital microscope (Keyence VHX 7000) to ensure the cuts are smooth and to check for particle contamination. Typical images are shown in Fig.~\ref{fig:Capillary Images}(a)-(b). The fiber ends are placed in a tubing sleeve to prevent breakage and to allow for simple attachment to the custom-built fiber adapters (similar to those used in Refs.~\citenum{Frosch2013,Yan2017}, technical drawings provided in Ref.~\citenum{mythesis}).

\begin{figure*}[!t]
\includegraphics[width=0.60\textwidth]{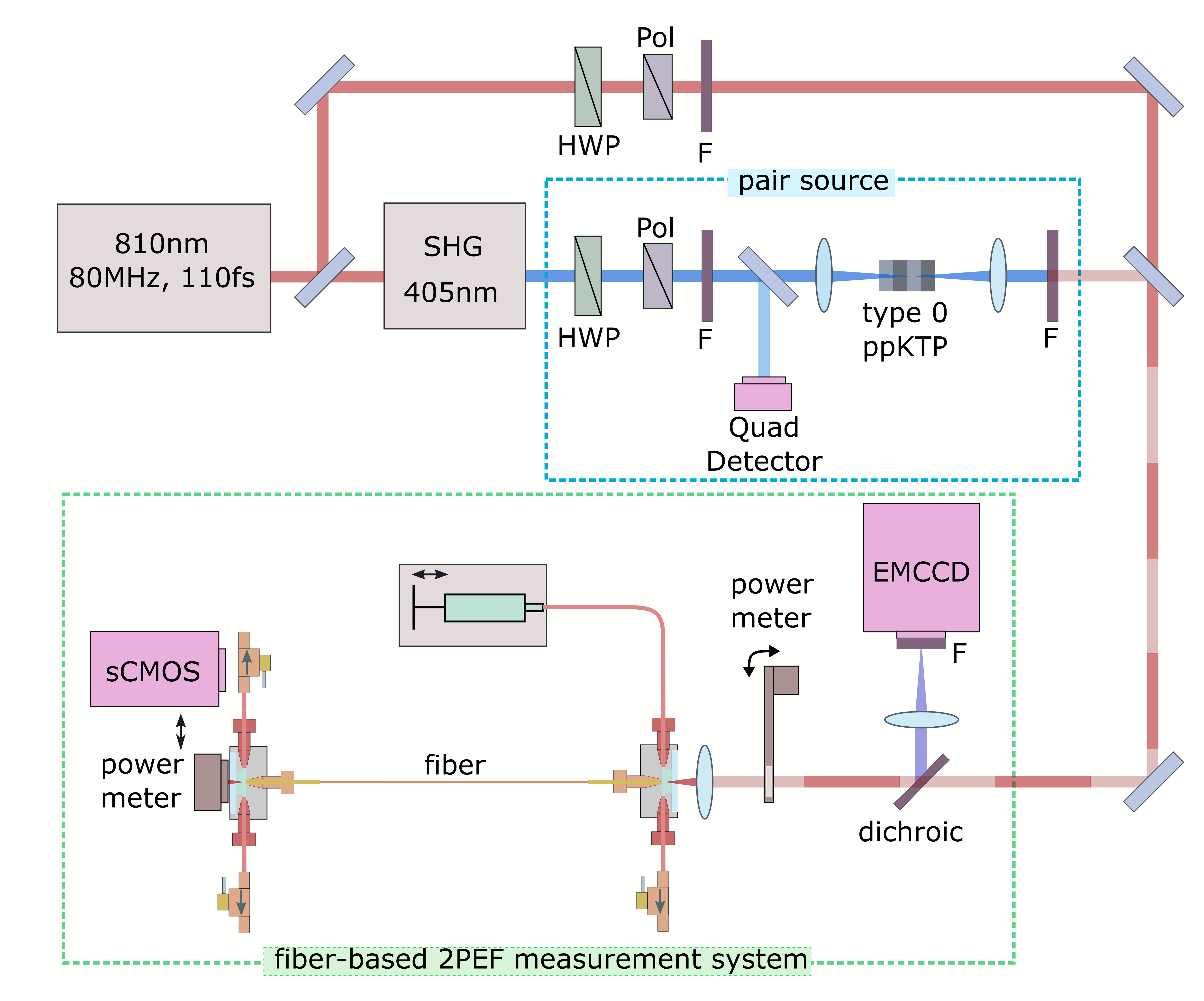}
\caption[Schematic of fluorescence experiment]{\label{fig:CapillarySetup}Schematic of the experimental setup. The 810~nm, 110~fs pulsed laser is split. Along one path the light is sent through a second-harmonic generation unit to frequency double the light. It's then sent through a telescope containing a type-0 ppKTP crystal used to generate photon pairs. Along the other path the beam is directed around the nonlinear crystals and is recombined with the light along the other path at a back-polished mirror. Either beam is sent through a dichroic beamsplitter and focused into the fiber held in two fiber adapters at either end. Solvent or a solution is pumped into the fiber using a syringe and syringe pump connected by PEEK tubing to one of the fiber adapters. Light transmitted through the fiber can be detected on a power meter or an sCMOS camera. A fraction of the fluorescence generated in the solution is directed out of the fiber in the direction of the dichroic beamsplitter, where it is reflected and focused onto the EMCCD camera. SHG = second-harmonic generation, HWP = half-wave plate, pol = polarizer, F = spectral filters, 2PEF = two-photon excited fluorescence}
\end{figure*}

The fiber and tubing sleeves are secured into the fiber adapters as shown in Fig.~\ref{fig:CapillarySetup}. One fiber adapter is connected to a syringe placed into a syringe pump that is used to fill the fiber. Both fiber adapters are connected to valves that serve as drainage ports. The fiber adapters are fitted for fused silica optical windows, which are sealed onto the front for coupling light into and out of fiber. These windows also serve as viewports to check whether fluid has flowed through the fiber.

The optical and microfluidic setup is illustrated in Fig.~\ref{fig:CapillarySetup}. Here we give a brief overview of the setup. A complete list of the components is given in Ref.~\citenum{mythesis}. A tunable femtosecond pulsed laser operated at 810~nm (110~fs pulses, $\approx 9$~nm bandwidth) is used to generate SPDC photons for E2PA measurements and its direct output is used for C2PA measurements. For photon pair generation, the output is first directed through a second harmonic generation unit to produce 405~nm light ($\approx 3$~nm bandwidth). The average power is controlled using a half-wave plate and a polarizer. Spectral filters are used to remove any remaining 810~nm light or unwanted harmonics. A small portion of this beam is picked off by a beam sampler to monitor the power and beam pointing stability of the laser on a quadrant (quad) detector. The blue light is focused ($f=300$~mm lens) into a 10~mm-long type-0 periodically-poled potassium titanyl phosphate (ppKTP) crystal to generate photon pairs centered at 810~nm. The crystal is temperature controlled at $40.00\pm0.01\,^{\circ}$C. The photon pairs are collimated with a 50~mm lens and any remaining pump light is filtered out. 

In our previous report\cite{Parzuchowski} we found that this source has a spectral bandwith of approximately 75~nm full width at half maximum (FWHM) under these operating conditions. We measure that this source produces $8.25\times10^9$~photons~s$^{-1}$~mW$^{-1}$ using a scientific Complementary Metal–Oxide–Semiconductor (sCMOS) camera placed directly after the crystal and collimating lens. This corresponds to a mean photon number of $\approx 5000$~photons~pulse$^{-1}$ at our maximum pump power of 49.2~mW. We can estimate the number of SPDC spatial modes based on the fraction of the total photon rate that is collected into the few-mode fiber and accounting for losses due to fiber coupling, and absorption and scattering in toluene as discussed in the Supporting Information. This yields $\approx740$ spatial modes, which serves as a lower bound since up to 16 spatial modes can propagate in our fiber. Using this estimate, we find that we generate $\approx 6.8$~photons~pulse$^{-1}$~spatial mode$^{-1}$. The entanglement time of the SPDC is estimated using a discrete Fourier transform of the joint spectral amplitude, which is estimated using the measured joint spectral intensity (shown in Ref.~\citenum{Parzuchowski}) and a phase estimated using the calculated group delay dispersion accumulated by the SPDC between the center of the crystal and its position along the fiber length. The broadening of the entanglement time across the length of the fiber is taken into account in our calculations (Eq.~\eqref{eq:Te} of Supporting Information). At the entrance of the fiber the entanglement time is 1070~fs. The entanglement area was not measured, but we estimate it to be in the range of 2.1~$\mu$m$^2$ to 18~$\mu$m$^2$. The lower bound is set by a diffraction limited spot size (see Ref.~\citenum{Parzuchowski}). The upper bound is set by the spot size of the fundamental mode of the fiber. 

For the C2PA measurements, the pulsed laser is directed around the nonlinear crystals and through a half-wave plate, polarizing beamsplitter and neutral density filters for control over the power. The light from both sources is recombined at a back-polished mirror. Either source is directed into the fiber-based two-photon excited fluorescence (2PEF) measurement system. First the light propagates through a dichroic beamsplitter and is focused ($f=10$~mm) into the $5$~$\mu$m-diameter-core fiber. Light that is transmitted through the 37~cm-long fiber can be detected on a power meter or imaged on an sCMOS camera as shown in Fig.~\ref{fig:Capillary Images}(c). The power before the fiber can be measured with a power meter that flips into the beam path, which doubles as a beam block for background measurements. Any fluorescence generated in the core of the fiber and guided out in the direction opposite of the 810~nm excitation beam is reflected at a dichroic beamsplitter and focused onto an Electron-Multiplying Charge-Coupled Device (EMCCD) camera as shown in Fig.~\ref{fig:Capillary Images}(d). Spectral filters are used to remove scattered 810~nm light. 

Two fibers are used for the experiments: fiber 1 is used for the two lower sample concentration C2PA measurements (experiments 1 and 2) and fiber 2 is used for the highest sample concentration C2PA measurement (experiment 3) and for the E2PA measurement. The summed fiber scattering and toluene absorption coefficients are measured using long exposure images taken by a smartphone camera. The integrated intensity along the length of the fibers is fit to an exponential decay function as shown in Fig.~\ref{fig:ScatterFiber}. The scattering coefficient at 810~nm is determined to be negligible whereas the absorption coefficient is equated to that found in literature\cite{Kedenburg2012}. At 458~nm (near the fluorescence maximum) the summed coefficient is measured to be 0.093~cm$^{-1}$ for fiber 1 and 0.034~cm$^{-1}$ for fiber 2. 

The measured transmission efficiency of the laser through the fiber was about $40\%$, $43\%$ and $47\%$ for experiments 1, 2 and 3 respectively; losses due to coupling, absorption in toluene ($11\%$) and scattering are all accounted for in this efficiency. This high transmission is consistent with the light primarily occupying the fundamental mode. The average laser transmission efficiency serves as a best estimate for the transmission efficiency of the fundamental mode of SPDC ($43\%$). If we consider all spatial modes, including those that are irrelevant in the estimation of E2PA signals since they are not coupled into fiber, we can estimate an effective transmission efficiency of $0.07\%$. The huge difference in transmission efficiency between a single mode and $\approx 740$ modes illustrates that filtering the SPDC to a small number of modes may give the same result as the experiment we present. The fiber has a lower retention of light in higher-order modes and thus it effectively acts like a single-mode filter. Details on the definition and measurement of transmission efficiency for both laser and SPDC are given in the Supporting Information.

To estimate the efficiency of coupling one photon into fiber given that its spatially-correlated partner photon is coupled into fiber, we estimate an effective Klyshko\cite{Klyshko} efficiency using SPDCalc\cite{SPDCalc}. SPDCalc uses various parameters of our pump beam, crystal, lenses and fiber, to calculate the overlap integral of three Gaussian modes---one for signal photons, one for idler photons, and one for the collected single mode in fiber---along the length of the crystal. This yields $\eta_K^{'}=0.94$. This differs from a measured Klyshko efficiency, $\eta_K$, because it does not account for any single photon loss between photon pair generation and collection into fiber. We multiply this value by the measured free-space transmission efficiency and the single-mode coupling efficiency to estimate  $\eta_K=0.25$. This value is used to estimate that $25\%$ of photons coupled into fiber are part of an intact pair. See Eq.~\eqref{eq:Klyshko} on Klyshko efficiency in the Supporting Information for more details.

\begin{figure*}[!t]
\includegraphics[width=0.99\textwidth,trim=0 0 0 0, clip]{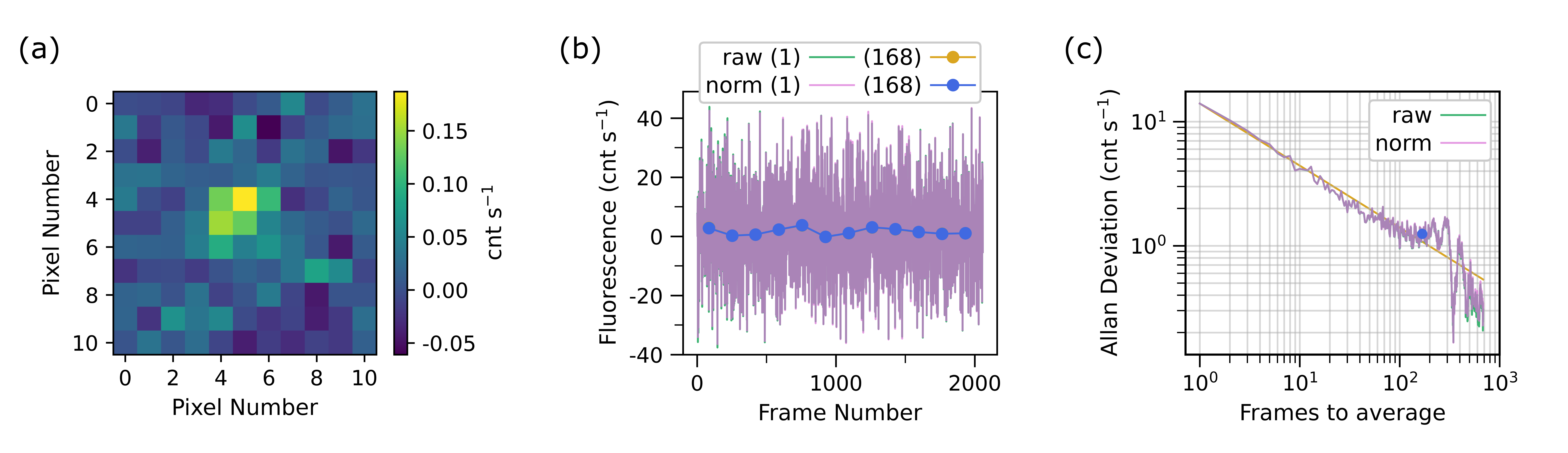}
\caption[Example dataset for a fluorescence measurement]{The data series for $1.75$~nW average laser excitation power used with the $2.30$~mM sample. (a) The average background-subtracted image is shown for the 11x11 superpixel region of interest. (b) The fluorescence rate calculated from each frame's background-subtracted image is plotted in teal for the raw data or pink for the normalized (norm) data as a function of frame number. The overlap of the raw and normalized data is indicated by the purple trace. Averaging in portions of 168 frames is shown in gold for the raw data or blue for the normalized data. Averaging the normalized data gives a photon rate of 1.6~counts per second (cnt~s$^{-1}$). The normalized data for both the individual and averaged data follows the raw data closely and thus nearly perfectly overlaps the raw data on the plot. This trend indicates a fairly stable laser. (c) The Allan deviation of the photon rate is plotted as a function of number of frames used to average. The teal data shows the Allan deviation of the raw data and the pink shows that for the normalized data. The overlap of the two traces is indicated in purple. A gold $1/\sqrt{N}$ (where $N$ is the number of frames) line is used to guide the eye. The Allan deviation value used in the analysis is indicated by a blue data point, corresponding to averaging portions of 168 frames with an Allan deviation of 1.3~cnt~s$^{-1}$. This data series consists of 2058 frames, which required about 13~hours to collect.}
\label{fig:exampleDataFiber}
\end{figure*}
The sample AF455 is chosen for our measurements due to its large C2PA cross-section at 810~nm and its solubility in toluene. The “AF455” fluorophore is provided by the Air Force Research Laboratory\cite{Kannan2004,Rogers2004}. The toluene used to prepare the sample has a purity of $\ge99.98\%$.  The concentrations were calculated from the one-photon absorption spectra by use of a UV-VIS-NIR spectrophotometer (Agilent Cary 5000 Scan). The concentration of the sample is monitored over the course of a measurement and found to vary by $\le12\%$

\section*{Results and Discussion}
\label{sec:fiberResults}
\textbf{Classical Two-Photon Absorption Measurements.} Classical two-photon excited fluorescence (C2PEF) datasets were gathered for three different concentrations of AF455 in toluene. Data series were acquired at a variety of laser powers until the uncertainty, set by the Allan deviation, reached a value at least 25\% smaller than the photon rate derived from the measurement. For high excitation powers, the acquisition time was as little as five minutes, which produced uncertainty values of less than 1\%. Our longest data series was acquired over a 13-hour period, which lowered the uncertainty to the point where the previously indistinguishable signal could be discerned.

An example data series is shown in Fig.~\ref{fig:exampleDataFiber} for the lowest average excitation power ($W_0$ in Eq.~\eqref{eq:avgPower} of the Supporting Information) of $1.75$~nW. This data series was acquired over $\approx 13$~hours during experiment 3. In Fig.~\ref{fig:exampleDataFiber}(a), the background-subtracted image, which is averaged over the duration of the data series, is shown for the 11x11 superpixel (24x24 pixels per superpixel) region of interest. Near the center of the image a resolved bright spot shows the signal. Without binning this image looks similar to Fig.~\ref{fig:Capillary Images}(d).
\begin{figure*}
\centering
\includegraphics[width=0.98\textwidth]{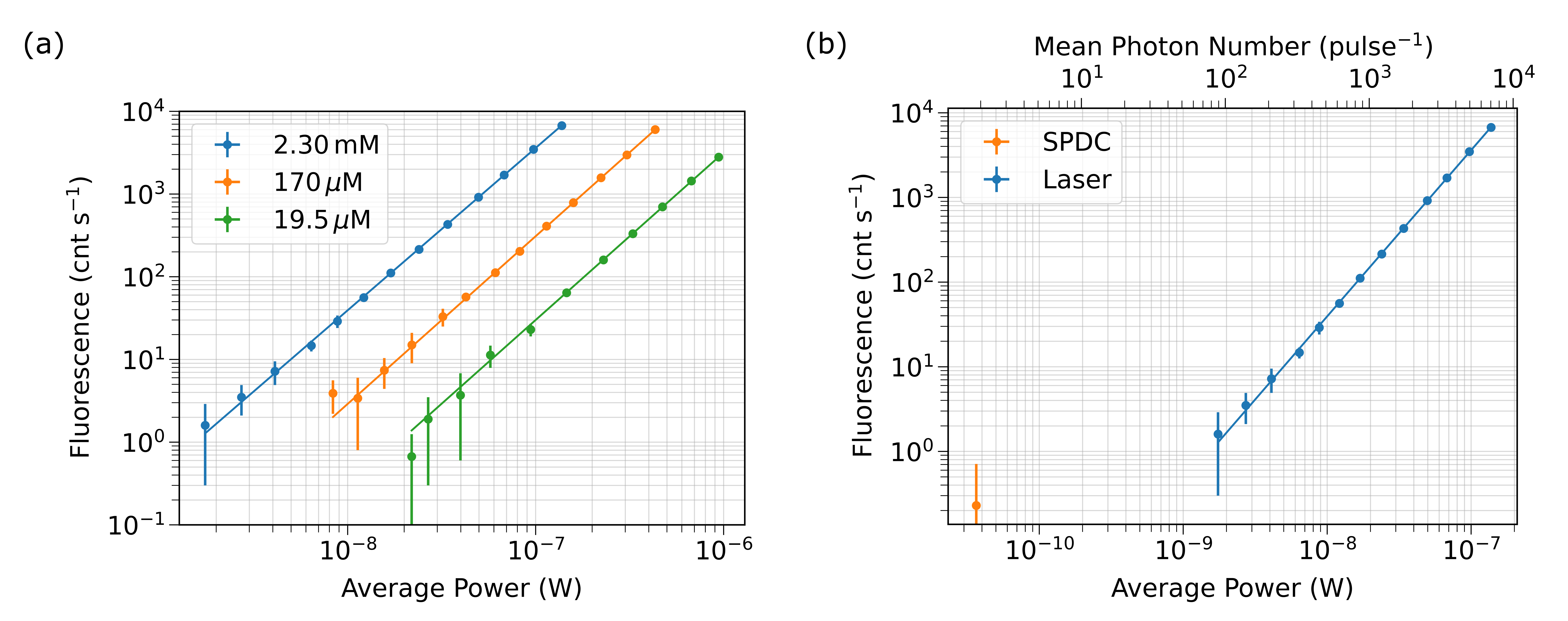}
\caption[Results from AF455 power dependence measurements]{\label{fig:CapillaryResults1}(a) The fluorescence rate measured as a function of the average laser power input to the fiber for a sample of $2.30$~mM, $170\,\mu$M and $19.5\,\mu$M AF455 in toluene shown in blue, orange and green respectively. The linear fits to the datasets have slopes of 1.96, 2.03 and 2.02 respectively. (b) The fluorescence rate measured as a function of the power input to the fiber for both SPDC (orange) and laser (blue) excitation for a sample of $2.30\,\mu$M AF455 in toluene. The SPDC measurement is performed at a power 48-fold lower than the minimum laser power and is indistinguishable from zero.}
\end{figure*}

In Fig.~\ref{fig:exampleDataFiber}(b) the fluorescence rate extracted from each frame's background-subtracted image is plotted as a function of frame number for both raw (teal) and normalized (pink) data. The normalized data accounts for any changes in laser excitation power (see Supporting Information). In Fig.~\ref{fig:exampleDataFiber}(c) the Allan deviation of the fluorescence rate is plotted for different numbers of frames ($N$) used for averaging for both raw (teal) and normalized (pink) data. For both Figs.~\ref{fig:exampleDataFiber}(b) and (c), the overlap of the raw and normalized data is indicated by a purple trace. The Allan deviation should follow a $1/\sqrt{N}$ line in the absence of noise sources at that combined frame rate, thus this gold line is used to pick a near optimum number of frames to average. Here, 168 frames is selected, which corresponds to an Allan deviation of 1.3~cnt~s$^{-1}$. We average the data in portions of 168 frames and plot the result in Fig.~\ref{fig:exampleDataFiber}(b) for the raw (gold) and normalized (blue) data. These measurements have an average value of 1.6~cnt~s$^{-1}$.

The results of our fluorescence measurements for three concentrations of AF455 in toluene are plotted in Fig.~\ref{fig:CapillaryResults1}(a), which shows the fluorescence count rate as a function of average excitation power. The fits to the datasets are linear regressions performed on a log-log scale. The slopes of the fits are all within $0.05$ of $2.00$, which confirms the two-photon origin of the signals. The lowest data point (acquired using the data shown in Fig.~\ref{fig:exampleDataFiber}) is measured for the highest concentration at a power of $1.75$~nW. The signals shown here are compared with our model using measured and calculated parameters in Fig.~\ref{fig:ConcNormFlu} and show qualitative agreement. 

For all three AF455 concentrations, we find that C2PA can be measured at excitation powers significantly lower than is typically possible with a free-space technique. Table~\ref{Tab:compare} compares the parameters used in this experiment to our previous experiment performed in free space\cite{Parzuchowski}. In our previous report, for a sample of $1.10$~mM AF455 in toluene, C2PA can be measured down to $79$~nW. In this work, we show a 45-fold improvement. The advantage of this platform is ultimately due to the capability to generate high photon fluxes with low average powers by focusing the light to a very small spot size and maintaining it over the length of the fiber. The peak photon flux at our minimum excitation power of $1.75$~nW is $1.1\times10^{22}$~photons~cm$^{-2}$s$^{-1}$; in the earlier technique at the minimum power of $79$~nW, the peak photon flux was $1.3\times10^{21}$~photons~cm$^{-2}$s$^{-1}$. Thus, an 8.5-fold higher photon flux was achieved with a 45-fold lower power.
\begin{figure*}[!t]
\includegraphics[width=0.99\textwidth,trim=0 0 0 0, clip]{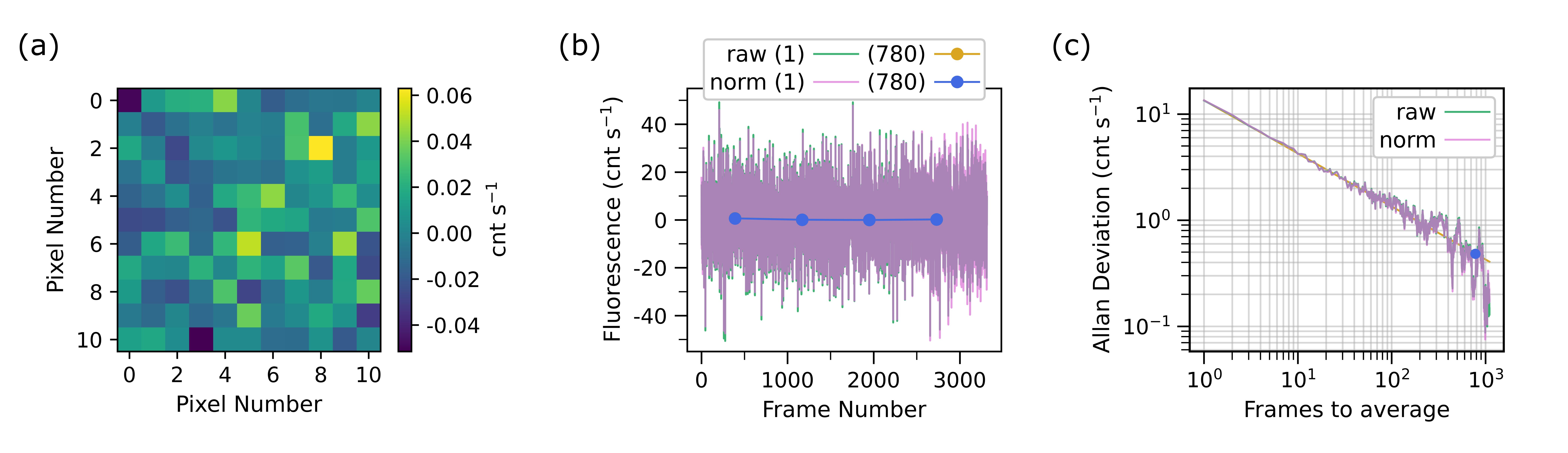}
\caption[SPDC dataset]{The data series for SPDC excitation of the $2.30$~mM sample. (a) The average background-subtracted image is shown for the 11x11 superpixel region of interest. (b) The fluorescence rate calculated from each frame's background-subtracted image is plotted in teal for the raw data or pink for the normalized (norm) data as a function of frame number. The overlap of the raw and normalized data is indicated by a purple trace. Averaging in portions of 780 frames is shown in gold for the raw data or blue for the normalized data. Averaging the raw and normalized data gives a photon rate of 0.23~cnt~s$^{-1}$ and 0.22~cnt~s$^{-1}$ respectively. The normalized data for both the individual and averaged data follows the raw data closely and thus nearly perfectly overlaps the raw data on the plot. This trend indicates a fairly stable laser. (c) The Allan deviation of the photon rate is plotted as a function of number of frames used to average. The teal data shows the Allan deviation of the raw data and the pink shows that for the normalized data. A $1/\sqrt{N}$ (where $N$ is the number of frames) line is used to guide the eye and is shown in gold. The Allan deviation value used in the analysis is indicated by a blue data point, corresponding to averaging portions of 780 frames with an Allan deviation of 0.48~cnt~s$^{-1}$ for both raw and normalized data. This data series consists of 3316 frames, which required about 24~hours to collect.}
\label{fig:SPDCdataset}
\end{figure*}

Using Eq.~\eqref{eq:C2PAxSectionFiber} of the Supporting Information, C2PA cross-sections are derived using the fits to each of the three concentration datasets. We derive values of ($570\pm190$)~GM, ($340\pm120$)~GM and ($250\pm80$)~GM for the $2.30$~mM, $170\,\mu$M and $19.5\,\mu$M samples, respectively. The average of these is ($390\pm80$)~GM which is in line with the cross-sections reported for AF455\cite{deReguardati2016,Parzuchowski}. 

\textbf{Entangled Two-Photon Absorption Measurement.} The SPDC source was used exclusively with the $2.30$~mM sample. Using the high concentration sample maximizes the likelihood of measuring entangled two-photon excited fluorescence (E2PEF). The results from the SPDC data series are shown in Fig.~\ref{fig:SPDCdataset}. For this measurement the pump power was nearly maximum for an average of 49.2~mW, which generates $4.06\times10^{11}$~photons~s$^{-1}$. A small number of modes of SPDC are effectively coupled into fiber, reducing the photon rate to $1.49\times10^{8}$~photons~s$^{-1}$. The average background-subtracted image is shown in Fig.~\ref{fig:SPDCdataset}(a), where there is no discernible bright spot due to a signal. In Fig.~\ref{fig:SPDCdataset}(c), the Allan deviation is plotted, which is used to determine a near optimal averaging of 780 frames. This averaging corresponds to an uncertainty of 0.48~cnt~s$^{-1}$ for both the raw (teal) and normalized (pink) data. In Fig.~\ref{fig:SPDCdataset}(b), the fluorescence rate from each frame is plotted as well as the averaged (in 780 frame portions) fluorescence rate. The average fluorescence rate for the raw (gold) data is ($0.23\pm0.48$)~cnt~s$^{-1}$ and for the normalized (blue) data is ($0.22\pm0.48$)~cnt~s$^{-1}$. The magnitude of the signal is not significantly different from zero, and therefore we conclude that there is no resolvable signal. 
\begin{table*}[t]
%    \vspace*{0.2cm}
    \caption{Comparison of fiber to free-space AF455 measurements.}
    \label{Tab:compare}
%    \vspace*{-0.2cm}
%    \centering
    \renewcommand{\tabcolsep}{14pt}
    \begin{tabular*}{\textwidth}{ llcc }
    \midrule
    Parameter & \multicolumn{1}{c}{unit} & Fiber (this work) & Free space (Ref.~\citenum{Parzuchowski}) \\ \midrule
    Concentration & mM  & 2.30 & 1.10  \\
    Min. laser power (flux) & nW (cm$^{-2}$s$^{-1}$) & 1.75 ($1.1\times10^{22}$)\textsuperscript{\emph{a}} & 79 ($1.3\times10^{21}$)  \\
    C2PA cross-section  & GM & $390\pm80$ & $660\pm180$ \\
    Entanglement area & $\mathrm{\mu m}^2$ & 2.1 to 18 & 2.1 to 13,700\\
    Entanglement time & fs & 1070\textsuperscript{\emph{a}} & 1620 \\
    SPDC power (flux) & pW (cm$^{-2}$s$^{-1}$) & 36.5 ($7.7\times10^{19}$)\textsuperscript{\emph{a}} & 2200 ($2.1\times10^{18}$)\\
    SPDC loss & \% & 73\textsuperscript{\emph{a}} & 24 \\
    Intact pairs & \% & 25\textsuperscript{\emph{a}} & 58 \\
    E2PA cross-section UB & cm$^2$ & $(5.8\pm2.3)\times10^{-24}$ & $(2.1\pm0.5)\times10^{-25}$ \\
    $R^{\mathrm{UB}}$ &  & \multicolumn{2}{c}{8.5} \\
    \midrule
    \end{tabular*}
    \\ \textsuperscript{\emph{a}} This value is valid at the front of the fiber ($z=0$).
\end{table*}

Although E2PEF was indistinguishable from zero, there are numerous conclusions to be drawn from this measurement. The C2PEF dataset for the $2.30$~mM sample is plotted alongside our E2PEF measurement in Fig.~\ref{fig:CapillaryResults1}(b). Here we show measured fluorescence count rates as a function of the average power of either the SPDC (orange) or laser (blue) source. For the E2PEF measurement, we couple about 36.5~pW ($\approx 1.9$~photons~pulse$^{-1}$) of SPDC into the fiber. This plot demonstrates the difference between the SPDC power and the minimum laser power, which are only 48-fold apart.

We can compare these E2PA results with those done in free space (Table~\ref{Tab:compare}). Notably, the power of SPDC at the sample is 60-fold lower in the present experiment. This is because the free-space experiment uses all the photons from $\approx 740$ spatial modes, whereas the fiber acts as a few-mode filter. Despite this lower power, the peak photon flux is $7.7\times10^{19}$~photons~cm$^{-2}$s$^{-1}$. In the free-space studies the SPDC peak photon flux was $2.1\times10^{18}$~photons~cm$^{-2}$s$^{-1}$, which is 37-fold lower. It should be noted that some of the advantage of the higher SPDC photon flux is hindered by the lower fraction of intact photon pairs propagating in the fiber, which is 25\% for fiber and 58\% for free space. The intact pair rate is accounted for using the estimated Klyshko efficiency in the calculations in the Supporting Information.

Using Eq.~\eqref{eq:E2PAxSectionUB} in the Supporting Information, we set an upper bound on the E2PA cross-section of this sample at $(5.8\pm2.3)\times10^{-24}$~cm$^2$. This upper bound is for an entanglement time of 1070~fs and an entanglement area in the range $2.1$~$\mu$m$^2$ to $18$~$\mu$m$^2$. This entanglement time is an estimated value for the SPDC at the entrance of the fiber adapter. The broadening of the entanglement time in the fiber is among numerous parameters accounted for in the calculation of the upper bound (see Supporting Information). 

We can compare the upper bound set here to the upper bound set in Ref.~\citenum{Parzuchowski} for the same sample as shown in Table~\ref{Tab:compare}. Since the E2PA cross-section is known to scale with parameters of the excitation beam that varied between the two experiments, we cannot compare the two numbers alone. For instance, in an experiment where the SPDC beam is tightly focused at the sample, the entanglement area typically decreases proportionally, leading to a larger E2PA cross-section compared to experiments with gentler focusing. Similary, if the SPDC is sent through dispersive elements, the temporal separation between photons in a pair broadens---especially for SPDC sources with broad bandwidths---resulting in an increased entanglement time and a reduced E2PA cross-section. To properly compare the upper bounds, we use the probabilistic model (see for example Ref.~\citenum{Parzuchowski}) and arrive at the ratio $R^{\mathrm{UB}}$,
\begin{equation}
R^{\mathrm{UB}} = \frac{\sigma_E^{\mathrm{UB},1}T_e^1A_e^1}{\sigma_E^{\mathrm{UB},2}T_e^2A_e^2},
\end{equation}
where $T_e$ and $A_e$ are the entanglement time and area, $1$ indicates the current experiment, and $2$ indicates the previous experiment. 

Although we do not know the exact entanglement areas of the SPDC used in either experiment, the same ppKTP crystal, pump laser and pump focusing conditions were used in both experiments. Assuming that the spatial correlations of the SPDC are maintained as it propagates through the optical system, the entanglement area should scale with the beam size at the image plane of the crystal. Since the optical system is designed to image the crystal into the sample, we can use this logic to estimate the ratio of the entanglement areas in the previous experiment to that in the current experiment as the ratio of the respective beam sizes of the SPDC at the sample position. For a fair comparison, we use the beam size of all $\approx 740$ spatial modes at the focus of the lens used for fiber coupling, which is $6,350$~$\mu$m$^2$. In free space, the beam size at its focus in the sample was $13,700$~$\mu$m$^2$. Plugging in the numbers we arrive at $R^{\mathrm{UB}}$ of 8.5. This result indicates that the current experiment sets an upper bound that is 8.5-fold larger than the previous experiment, and is thus 8.5-fold less stringent.

\section*{Conclusions}
We presented a toluene-filled fiber platform for two-photon excited fluorescence measurements. We used this platform to measure classical two-photon absorption of various concentrations of the sample AF455. The results of the C2PA measurements show that 2PA can be measured at extremely low powers ($1.75$~nW) using an excitation volume of only 7.3~nL. This power is 45-fold lower than we achieved previously in free-space under similar conditions (Ref.~\citenum{Parzuchowski}). Furthermore, we derived a classical two-photon absorption cross-section for AF455 of (390$\pm$80)~GM. These results emphasize the advantages of liquid-core-fiber platforms for sensing applications where one could instead use higher powers to sense very low concentrations of analytes. 

With photon pair excitation, we saw no evidence of a signal, and set an upper bound on the E2PA cross-section of AF455. Our calculation of $R^{\mathrm{UB}}$, which uses $\sigma_E^{\mathrm{UB}}$ to account for many experimental factors such as loss and the broadening of the entanglement time, shows that the bound we set on the entangled two-photon absorption cross-section is 8.5-fold larger than that set in Ref.~\citenum{Parzuchowski}. Thus, this technique does not provide a higher E2PA sensitivity compared to the free-space technique. However, since the SPDC loss reduces the upper bound in a quadratic manner, efforts to decrease the loss would greatly improve the sensitivity of this technique. This work furthers the growing body of research that has presented null results when attempting to measure entangled two-photon absorption (Refs.~\citenum{Mikhaylov2020,Parzuchowski,Mazurek,Landes2021,Mikhaylov2022,CoronaAquino2022,Hickam2022,Arango2023,Gabler2023,Qian2024,He2024,Landes2024}).

\begin{acknowledgement}
This work was supported by NIST and by the NSF Physics Frontier Center at JILA (PHY 2317149) and by the NSF-STROBE center (DMR 1548924). We thank L. K. Shalm, S. Mukherjee, A. McLean, H. Greene, K. Thatcher, J. Sipe and C. Drago for valuable suggestions and discussions on the experimental design. We are grateful to T. Loon-Seng Tan and T. Cooper for providing molecular samples. Certain commercial equipment, instruments, or materials are identified in this paper in order to specify the experimental procedure adequately. Such identification is not intended to imply recommendation or endorsement by NIST, nor is it intended to imply that the materials or equipment identified are necessarily the best available for the purpose.

\end{acknowledgement}

%% suppinfo environment
%%%%%%%%%%%%%%%%%%%%%%%%%%%%%%%%%%%%%%%%%%%%%%%%%%%%%%%%%%%%%%%%%%%%%
\section*{Supporting Information}

\section*{Experimental Characterization}
\label{Sec:FiberChar}
In this section we discuss the experimental characterizations we performed to derive C2PA cross-sections using Eq.~\eqref{eq:C2PAxSectionFiber} and an E2PA cross-section upper bound using Eq.~\eqref{eq:E2PAxSectionUB}. First, we discuss the transmission efficiency and propagation losses of the fiber. Next, we discuss the considerations needed to switch from single-mode laser light to multimode SPDC. We characterize the number of spatial modes and the Klyshko efficiency. Afterwards, we discuss images of the excitation and fluorescence light at the fiber face for qualitative identification of the mode content of the fiber. Finally, we discuss our characterization of the setup's dispersion.
\begin{figure*}[!tbh]
\centering
\includegraphics[width=0.99\textwidth,trim=10 10 10 10, clip]{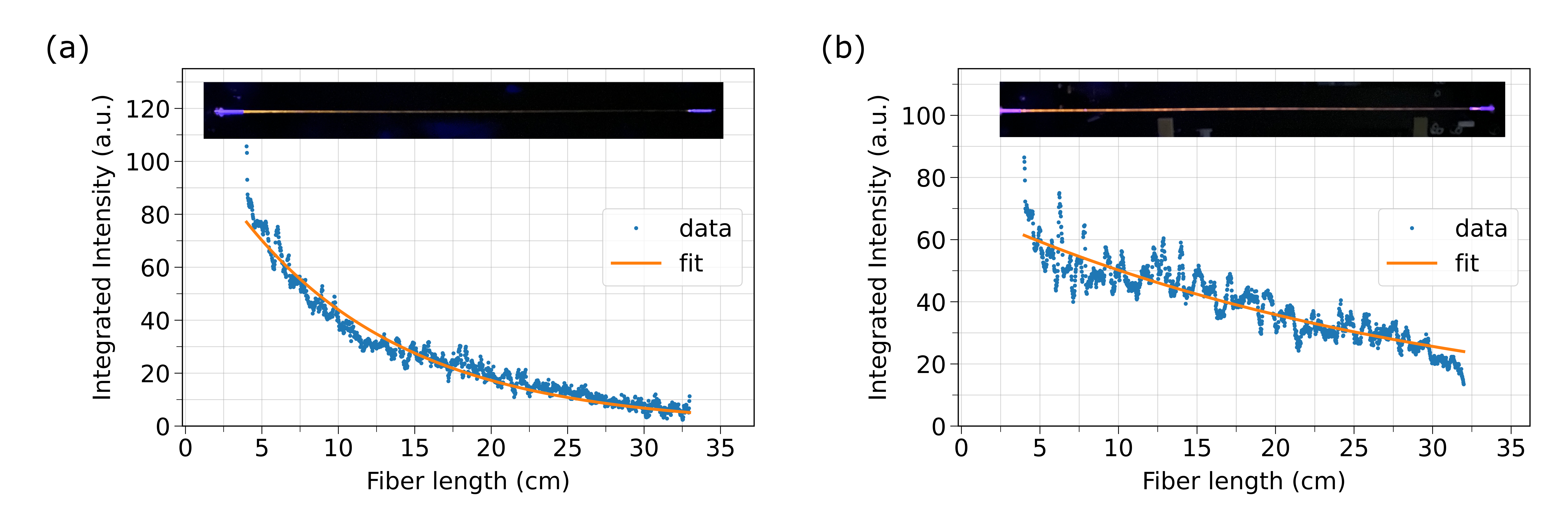}
\caption[Scatter of alignment laser along fiber length]{\label{fig:ScatterFiber}Images of the scatter from the alignment laser (inset) with corresponding plots of the integrated intensity (blue) of the scatter as a function of fiber length for (a) fiber 1 and (b) fiber 2. The intensity of the scatter is higher at the ends due to secondary reflections off the fiber tubing sleeves and fittings. These regions are removed from the plotted data. The fiber is not visible inside of the fittings and fiber adapters. The integrated intensity is fit (orange) to an exponential decay.}
\end{figure*}

Before C2PA experiments we maximize the transmission of the 810~nm laser light through the fiber. We measure the power before and after the fiber using power meters, and the ratio of the power transmitted out $W_\mathrm{out}$ (W) to the incident power $W_\mathrm{in}$ (W) is what we call our transmission efficiency,
\begin{eqnarray}
    \eta_T &=& \frac{W_\mathrm{out}}{W_{\mathrm{in}}} \\ 
    &=& \eta_C \eta_A(\lambda=\lambda_e,z=l) \eta_S(\lambda=\lambda_e,z=l). \nonumber
\end{eqnarray}
This efficiency is a product of coupling ($\eta_C$), absorption ($\eta_A$) and scattering ($\eta_S$) efficiencies. The latter two efficiencies are evaluated at the excitation wavelength ($\lambda_e=810$~nm) and for the entire length of the fiber $l$ (cm). The best alignment resulted in $\eta_T > 55\%$. However, during our measurements $\eta_T \approx 43\%$ but varied from data series to data series. This value for $\eta_T$ serves as a best estimate for the transmission efficiency for any aligned single spatial mode source at the same frequency, including a single mode of SPDC. Similarly, the values of the components of $\eta_T$: $\eta_C$, $\eta_A$ and $\eta_S$, derived using this measurement and the scattering measurements as described below, also serve as a best estimate of these parameters for any single mode of the same frequency. 

We characterize the loss of light along the length $z$ of the fiber by imaging the scattering. The intensity of the scattering is proportional to the power $W(z)$ (W), and thus should exponentially decay according to
\begin{eqnarray}
\label{eq:avgPower}
    W(z) = W_0 \times \\
    \textrm{Exp}\left(-(a_\mathrm{sol} (\lambda)+\epsilon_\mathrm{sam}(\lambda)c+\mu (\lambda))z\right), \nonumber
\end{eqnarray}
where $W_0$ (W) is the average power at $z=0$, $\lambda$ (nm) is the wavelength of the light, $a_{\mathrm{sol}}(\lambda)$ (cm$^{-1}$) is the absorption coefficient of the solvent, $\epsilon_{\mathrm{sam}}(\lambda)$ (~M$^{-1}$~cm$^{-1}$) is the extinction coefficient of the sample, $c$ (M) is the concentration of the sample and $\mu(\lambda)$ (cm$^{-1}$) is the scattering coefficient of the fiber. The parameter $W_0$ is related to the power measured before the fiber $W_{\mathrm{in}}$ (W) by $W_0 = \eta_C W_{\mathrm{in}}$. The power measured at the output of the fiber $W_{\mathrm{out}}$ is a good estimate for $W(l)$. The exponential terms related to absorption of the light are $\eta_A(\lambda,z)$ and to scattering of the light is $\eta_S(\lambda,z)$, thus we can rewrite Eq.~\eqref{eq:avgPower} as
\begin{equation}
 W(z) = W_{\mathrm{in}} \eta_C \eta_A(\lambda,z) \eta_S(\lambda,z).
\end{equation}

With only the solvent in the fiber ($c=0$), we send a 458~nm (near the peak of our sample's emission spectrum) alignment laser through the fiber and image the scattering on a smart phone camera using long exposure settings as shown in Fig.~\ref{fig:ScatterFiber}. The image is integrated along the width of the fiber (vertically) to find an integrated intensity as a function of fiber length. The intensity is fit to the exponential in Eq.~\eqref{eq:avgPower} to derive $a_\mathrm{sol}(458\,\mathrm{nm})+ \mu (458\,\mathrm{nm})=0.093$~cm$^{-1}$ for fiber 1 (Fig.~\ref{fig:ScatterFiber}(a)) and $a_\mathrm{sol}(458\,\mathrm{nm})+ \mu (458\,\mathrm{nm})=0.034$~cm$^{-1}$ for fiber 2 (Fig.~\ref{fig:ScatterFiber}(b)). For 458~nm fluorescence, traveling the entire length of the fiber corresponds to a loss of 96.8\% or 70.6\% of the light for fibers 1 and 2 respectively. We take the same measurement with 750~nm (a visible wavelength close to our excitation wavelength) light and observe no evidence of decay of the scatter along the length of the fiber. We estimate that $a_\mathrm{sol}(750\,\mathrm{nm})+\mu (750\,\mathrm{nm})$ is not large enough to measure this way. As mentioned in the main text, the values of $a_\mathrm{sol}(\lambda)$ are known. From literature,\cite{Kedenburg2012} $a_\mathrm{sol}(750\,$nm$)=0.0036$~cm$^{-1}$ and $a_\mathrm{sol}(810\,$nm$)=0.0030$~cm$^{-1}$. If we set $\mu (750\,\mathrm{nm})=\mu (810\,\mathrm{nm})=0$, we find that these absorption coefficients correspond to a loss of 12.5\% and 10.5\%, respectively, of the light along the length of the fiber. Considering with our best alignment $\eta_T>55$\%, and that some fraction of that light is lost due to the coupling efficiency in addition to the 10.5\% absorption loss, we estimate $\mu (810\,\mathrm{nm})\approx0$. This low scattering loss is consistent with the light primarily occupying the fundamental mode of the fiber. From here we can derive that for $\eta_T=43\%$, $\eta_C=48\%$. 

For SPDC, we are not able to measure $\eta_S$ due to the low photon rate, and thus cannot quantify the division of losses between $\eta_C$ and $\eta_S$. However, we can consider that $\eta_C$ and $\eta_S$ will likely only differ from that for laser light if the mode content in the fiber is different. If the SPDC does occupy more higher-order modes than laser light, the higher-order modes tend to have high $\eta_S$. This would cause a quick depletion of the modes from the fiber leading to minimal contributions to an E2PA signal. Furthermore, higher-order modes have spatial profiles that are less ideal for 2PA. Thus without losing the bulk of the proper physical model, we will operate under the assumption that only the fundamental mode is relevant in the E2PA measurement.

We followed a similar procedure to that used for measuring the transmission efficiency of the laser in order to estimate the number of SPDC spatial modes. Here we write the transmission efficiency as,
\begin{eqnarray}
\label{eq:SPDCtransEff}
    \eta_T &=& \frac{Q_\mathrm{out}M}{Q_{\mathrm{in}}^{\mathrm{mm}}} = \frac{Q_\mathrm{out}}{Q_{\mathrm{in}}}\\ 
    &=& \eta_C \eta_A(\lambda=\lambda_e,z=l) \eta_S(\lambda=\lambda_e,z=l). \nonumber
\end{eqnarray}
where $M$ is the total number of spatial modes of the incident light, $Q_\mathrm{in}$ (photons~s$^{-1}$) and $Q_{\mathrm{out}}$ (photons~s$^{-1}$) are the incident and output photon count rate and $Q_{\mathrm{in}}^{\mathrm{mm}}$ (photons~s$^{-1}$) is the incident photon count rates measured for a multimode (mm) light source. Unless specified with the labeling ``mm", all other quantities are for a single spatial mode. The values of $Q_{\mathrm{out}}$ and $Q_{\mathrm{in}}^{\mathrm{mm}}$ can be measured directly using an sCMOS camera. The incident photon count rate is an estimate for the incident photon count rate of the single mode coupled into fiber. The highest value achieved for the ratio of output photon count rate to incident multimode photon count rate is  $Q_\mathrm{out}/Q_{\mathrm{in}}^{\mathrm{mm}} \approx 0.058\%$. We can plug in the values of $\eta_S$, $\eta_A$ and $\eta_C$ determined using a laser source, as well as $Q_\mathrm{out}/Q_{\mathrm{in}}^{\mathrm{mm}}$ into Eq.~\eqref{eq:SPDCtransEff} to estimate $M\approx740$. This estimate for the number of spatial modes serves as a lower bound since up to 16 modes can be coupled into fiber as calculated using Eq.~\eqref{eq:Vnum} in the main text.

Thus far we have discussed coupling efficiency $\eta_C$ defined as the ratio of single photons incident on the fiber to single photons output from the fiber---all of which originate from the same single mode. This is in contrast to the Klyshko efficiency\cite{Klyshko}---a quantity typically measured for photon pair sources---defined as the probability of detecting a photon conditioned on the successful detection of its partner photon. For this work, we use a theoretical value for the Klyshko efficiency to aid in our estimation of intact photon pair rate in the fiber, $Q_{\mathrm{pairs}}$ (photon pairs~s$^{-1}$). We input various parameters of our pump beam, crystal, lenses and fiber into SPDCalc\cite{SPDCalc} which calculates the overlap integral of three Gaussian spatial modes---one for signal photons, one for idler photons, and one for the collected single mode in fiber---along the length of the crystal and estimates $\eta^{'}_{K}=0.94$. This differs from a measured Klyshko efficiency, $\eta_K$, because it does not account for any single photon loss between photon pair generation and collection into fiber. We can relate the two by,
\begin{eqnarray}
\label{eq:Klyshko}
\eta_K &=& \eta^{'}_{K} \eta_{F} \eta_{C} \\ 
       &=& \frac{2 Q_{\mathrm{pairs}}(z=0)}{Q_{\mathrm{singles}}(z=0)}, \nonumber
\end{eqnarray}
where $\eta_F$ is the free-space transmission efficiency between the center of the crystal to the fiber and $Q_{\mathrm{pairs}}(z)$ and $Q_{\mathrm{singles}}(z)$ (photons~s$^{-1}$) are the rates of photon pairs and single photons, respectively, in fiber. In later sections, we will refer to $Q_{\mathrm{singles}}(z)$ simply as $Q(z)$. The factor of 2 is used to align with the standard definition of $\eta_K$, which is typically measured using two fibers. In writing this equation and Eq.~\eqref{eq:SPDCtransEff}, we have assumed that the efficiencies of the various frequency modes are equal. We find that $\eta_K=0.25$.

We image the $810$~nm laser light at the output of the fiber using two lenses before the sCMOS camera, as shown in Fig.~\ref{fig:Capillary Images}(c). In this image, the bright larger and smaller concentric circles are the imaged fiber cladding outer diameter and the outline of the core modes, respectively. The larger offset circular shape is likely formed by light guided through the inside of the tubing sleeve as can be seen by comparison to the digital microscope image in Fig.~\ref{fig:Capillary Images}(b). Thus, we find that the light occupies some lower intensity cladding and tubing sleeve modes in addition to the dominant core modes.

The fluorescence is collected at the front of the fiber using the lens optimized to focus 810~nm light into the fiber, which roughly collimates the visible fluorescence. The light is reflected at the dichroic beamsplitter, and another lens focuses the image of the fiber face onto the EMCCD, as shown in Fig.~\ref{fig:Capillary Images}(d). In this image, the high intensity bright spot is from fluorescence guided through the core. Surrounding that spot is darkness likely from the cladding of the fiber, indicating that there is little light propagating through the cladding of the fiber. Radially outward from the cladding, light forms an offset circular shape and is likely from light guided through the core of the tubing sleeve as can be seen by comparison to the digital microscope image in Fig.~\ref{fig:Capillary Images}(b). A larger circular ring of light surrounds that core, and is likely from light guided through the ``cladding" of the tubing sleeve. Thus, the spatial mode content of the fluorescence in the fiber is more complex than Eq.~\eqref{eq:Vnum} in the main text assumes, however the highest intensity mode appears to be the fundamental mode of the fiber. 

To measure the free-space group delay dispersion (GDD), $D_0$ (fs$^2$), accumulated by the laser pulse as it propagates through the optical setup, we use the GDD tuning function of the laser. We increase the dispersion compensation in increments of 500~fs$^2$ and take C2PA measurements at each step. The step with the maximum C2PA signal corresponds to the value at which the GDD is optimally compensated at the input of the fiber. We measure $D_0 \approx 2000$~fs$^2$, $D_0 \approx 2000$~fs$^2$ and $D_0 \approx 4000$~fs$^2$ for experiments 1, 2 and 3 respectively. For experiments 2 and 3 the free-space dispersion is compensated for using the internal tuning of the laser system. To estimate $D_0$ accumulated by the SPDC as it propagates through the optical setup, we calculate the GDD of each optical element in its path from the center of the crystal to the entrance of the fiber. We estimate $D_0 \approx 2100$~fs$^2$. To determine the fiber group velocity dispersion (GVD) $\beta$ (fs$^2$~cm$^{-1}$), a COMSOL simulation is used to model the filled fiber. The low-order modes are solved for. For the fundamental mode, $\beta$ is identical to $\beta$ of toluene, thus there are no contributions from waveguide dispersion. However, for higher-order modes it is likely that $\beta$ is affected by waveguide dispersion. These higher-order modes are not considered in the calculations of the C2PA cross-section and E2PA cross-section upper bound, and $\beta$ of the fundamental mode is used.

\section*{Data Acquisition}
\label{Sec:DataAcq}
In the section we discuss the camera settings used for data acquistion, the characterization of camera baseline and dark count rates, the operations that go into measuring a single frame of data and the conversion of a camera signal to a detected fluorescence rate.

To determine the optimal camera settings, we wrote a script to calculate the expected signal to noise ratio (SNR) of our measurements at various EMCCD settings based on our expected signal levels. We determined the optimum camera settings to be: electron multiplying (EM) output amplifier, EM gain set to 30, preamplifier set to 1, 1~MHz horizontal shift rate, 10~s integration time and 24x24 pixels binned into a superpixel. Although the integration time could be increased further to increase the SNR, we found that clock-induced charge (CIC) occurred more frequently at those integration times. Any obvious CIC was removed from the data, which resulted in a removal of about 6.8$\%$ of the frames. To speed up frame readout on the EMCCD, a pixel region of interest (ROI) is selected and only those pixels are read out. To select an ROI, the fluorescence was imaged at a relatively high excitation power and the region with significant photon counts was selected. 

Before data series are acquired, the baseline and dark count rates of the camera are characterized. Both are characterized with the built-in camera shutter closed. The baseline data series are taken at the minimum integration time of the camera. The dark rate data series are taken at the 10~s integration time used for all 2PA data series. From each of these characterization data series we calculate an average value and uncertainty using an Allan deviation analysis like that shown in Fig.~\ref{fig:exampleDataFiber} for the fluorescence rate. An example baseline average in our ROI yields $560.4\pm1.3$~Analog to Digital Units per pixel (ADU~pixel$^{-1}$) and an example dark rate average in our ROI yields $2.66\pm0.06$~electrons~s$^{-1}$~pixel$^{-1}$. 

Each two-photon excited fluorescence (2PEF) data frame consists of a background and a signal measurement. The low profile power sensor (Thorlabs S130c) (Fig.~\ref{fig:CapillarySetup}) is flipped into the beam path to block the beam during a background measurement. For the signal measurement, the power sensor is removed from the beam. We ensured that the power sensor position did not affect the background signal with the laser on. The power sensor after the fiber is used to measure the laser power ($W_\mathrm{out}$) before each signal frame and is used to determine $W_0$ for plotting the power dependence of the signal (Fig.~\ref{fig:CapillaryResults1}(a)). For SPDC data frames, the sCMOS camera is placed at the output of the fiber and used to measure the SPDC power ($W_\mathrm{out}$) before each signal frame. The number of data frames acquired at each power varied. As the power decreased, the signal decreased and thus we acquired data for longer to lower the uncertainty. 

To calculate the fluorescence rate $F$ in cnt~s$^{-1}$ detected by the camera, the pixels of interest are integrated over to determine the camera signal $N$ (ADU). We converted $N$ to a count rate using,
\begin{equation}
    F = \frac{N S}{G T},
\end{equation}
where $S$ (electrons~ADU$^{-1}$) is the CCD sensitivity for the selected output amplifier and preamplifier, $G$ (electrons~cnt$^{-1}$) is the EM gain and $T$ (s) is the integration time. 

\section*{Data Normalization}
\label{Sec:DataNorm}
In this section we discuss the data normalization used to account for fluctuations in excitation power. For this, we made use of $W_\mathrm{out}$, which is measured for each frame, since it accounts for potential changes in $\eta_C$ due to laser pointing drifts unlike $W_\mathrm{in}$. 

For laser excitation, each frame's fluorescence rate $F_{i,\mathrm{raw}}$ (cnt~s$^{-1}$) was adjusted to a normalized fluorescence rate for frame $i$, $F_{i,\mathrm{norm}}$ (cnt~s$^{-1}$), using
\begin{equation}
F_{i,\mathrm{norm}} = F_{i,\mathrm{raw}}\frac{W^2_{\mathrm{out,avg}}}{W^2_{\mathrm{out,i}}},
\end{equation}
where $W_{\mathrm{out,i}}$ (W) is the power transmitted out of the fiber for frame $i$ and $W_{\mathrm{out,avg}}$ (W) is the average power transmitted out of the fiber for the entire data series. Here the normalization uses a quadratic power dependence in accordance with the power scaling of C2PA.

For SPDC excitation, each frame's fluorescence rate $F_{i,\mathrm{raw}}$ was adjusted to a normalized fluorescence rate for frame $i$ using
\begin{equation}
F_{i,\mathrm{norm}} = F_{i,\mathrm{raw}}\frac{W_{\mathrm{out,avg}}}{W_{\mathrm{out,i}}}.
\end{equation}
Here the normalization uses a linear power dependence in accordance with the expected power scaling of E2PA at sufficiently low power.

For all data series, the normalization has minimal effect and alters the value of the data points by less (typically much less) than one Allan deviation. This indicates that the laser is fairly stable throughout each measurement. The normalization is still used to demonstrate that a high level of characterization and analysis is essential, especially for the SPDC measurement.

For the SPDC measurement, a quadrant detector is used to monitor the pump power while the sCMOS camera is used to measure $W_\mathrm{out}$. The output power varied over a range of 24\% of the maximum value, and decreased monotonically relative to the pump power by 24\% over the course of the 24 hour measurement. The latter value is directly related to a decrease in coupling efficiency. The value $Q_\mathrm{out}/Q_{\mathrm{in}}^{\mathrm{mm}} = 0.058\%$ is the average value over the course of the measurement. Similarly, $\eta_C$, which is measured using the laser, is the average from three datasets each measured over the course of more than 24 hours and was thus also subject to drifts in coupling efficiency.

\section*{Calculating a C2PA Cross-Section}
\label{Sec:FibC2PAXSection}
In this section we discuss the assumptions and equations used to calculate a C2PEF signal. We arrive at an equation to derive a C2PA cross-section. Finally, we use these equations to model the concentration dependence of the signal and compare with our results.

In a similar manner to that shown in Ref.~\citenum{Parzuchowski}, we can model the C2PEF signal detected on the camera, $F_C$ (cnt~s$^{-1}$), as
\begin{equation}
\label{eq:FluRateFiber}
    F_C = g \int_0^l N_C(z) \int_{\lambda_1}^{\lambda_2} \gamma(z,\lambda) \kappa(\lambda) \Phi(\lambda) d\lambda dz ,
\end{equation}
where $g$ (pulses s$^{-1}$) is the pulse repetition rate, $l$ (cm) is the length of the fiber, $N_C(z)$ (excitations~cm$^{-1}$~pulse$^{-1}$) is the number of excitations per infinitesimal length of fiber $dz$ (cm) per pulse, $\lambda_{1}$ and $\lambda_{2}$ (nm) are wavelengths chosen such that the integral extends over the entire emission spectrum of the sample, $\gamma(z,\lambda)$ (cnt~photon$^{-1}$) is the component transmission efficiency, $\kappa(\lambda)$ is the geometrical collection efficiency and $\Phi(\lambda)$ (photon~excitation$^{-1}$~nm$^{-1}$) is the differential fluorescence quantum yield. A proper normalization of quantum yield is used such that $\Phi = \int_0^{\infty}\Phi(\lambda)d\lambda$ gives the value published in literature for the total quantum yield of the fluorophore. 

From Eq.~\eqref{eq:FluRateFiber}, we find that it is advantageous to increase the length of the fiber for all $z$ such that $N_C(z)\gamma(z,\lambda) >0$. This product never drops below zero, but can be zero if the photons are dispersed in time far enough that two photons are never temporally overlapped at that position $z$, or if the photon loss along the length $l$ is $100\%$ for either the excitation photons or the fluorescence photons, or some combination of both leading to a negligible product  $N_C(z)\gamma(z,\lambda) \approx0$. Since extra fiber length does not decrease the C2PEF signal, we try to make $l$ long enough that $N_C(l)\gamma(l,\lambda) \approx0$ so that we can achieve the highest signal at a given photon flux.

Here we define $\kappa(\lambda)$ as the fraction of fluorescence that can be collected by the fiber and directed out in the direction of the detector. We assume that the fluorescence is emitted isotropically and that the fraction of that light collected can be described by the solid angle of a cone with apex angle $2(90^{\circ}-\theta_c)$, where $\theta_c$ is the critical angle (see main text). Then we can write,
\begin{eqnarray}
     & \kappa (\lambda) = \frac{1}{2} \times \\
     & \left [ 1 - \textrm{cos}\left(\textrm{sin}^{-1}\left(\frac{\sqrt{n_\mathrm{core}^2(\lambda)-n_\mathrm{clad}^2(\lambda)}}{n_\mathrm{core}(\lambda)}\right)\right)\right ], \nonumber
\end{eqnarray}
where $n_\mathrm{core}(\lambda)$ and $n_\mathrm{clad}(\lambda)$ are the indices of refraction of the core and cladding materials (toluene and silica) at the wavelength of the fluorescence (peak wavelength $\lambda_f=451$~nm). 
\begin{table}[!b]
%    \vspace*{0.2cm}
    \caption{Summary of fixed parameters.}
    \label{Tab:fixedParam}
%    \vspace*{-0.2cm}
%    \centering
    \begin{tabular*}{\columnwidth}{ llcc }
    \midrule
    Parameter & \multicolumn{1}{c}{unit} & Laser & SPDC \\ \midrule
    $g$ & pulses~s$^{-1}$ & \multicolumn{2}{c}{$8\times10^7$} \\
    $h\nu$ & J & \multicolumn{2}{c}{$2.45\times10^{-19}$} \\
    $d_0$ & $\mu$m & \multicolumn{2}{c}{2.42} \\
    $a_{\mathrm{sol}}(\lambda_e)$\textsuperscript{\emph{a}} & cm$^{-1}$ & \multicolumn{2}{c}{0.003} \\
    $\tau_0$ & fs & $110$ & N/A \\
    $\beta$ & fs$^2$~cm$^{-1}$ & \multicolumn{2}{c}{$1034$} \\
    $\epsilon(\lambda_f)$\textsuperscript{\emph{a}} & M$^{-1}$~cm$^{-1}$ & \multicolumn{2}{c}{4417} \\
    $\kappa(\lambda_f)$\textsuperscript{\emph{a}} & & \multicolumn{2}{c}{0.0146} \\
    $\Phi$ & phot~excit$^{-1}$ & \multicolumn{2}{c}{0.67~[\hspace*{-4px}\citenum{Rogers2004}]} \\
    $F^{\mathrm{LB}}$ & cnt~s$^{-1}$ & \multicolumn{2}{c}{1.00} \\
    $\eta_F$ & & N/A & 0.565 \\
    $T_{e,0}$ & fs & N/A & $260$\\
    $S_0$ & & N/A & $2145$\\
    $Q(0)$ & photons~s$^{-1}$ & N/A & $1.49\times10^{8}$ \\
    $M$ & & 1 & 740 \\
    $\eta_K$ & & N/A & 0.25 \\
    \midrule
    \end{tabular*}
    \\ \textsuperscript{\emph{a}} $\lambda_e=810$~nm, $\lambda_f=451$~nm 
\end{table}

The component transmission efficiency describes the transmission efficiency of fluorescence from its point of generation $z$ (cm) to the detector, under the assumption that it is directed out of the fiber. It takes into account all the loss mechanisms from various optical components and the media the light propagates through. We define this quantity as
\begin{eqnarray}
\label{eq:compTransEff}
    \gamma(z,\lambda) &=& \eta_A(\lambda,z)\eta_S(\lambda,z)\mathcal{T}_\mathrm{w}(\lambda) \mathcal{T}_\mathrm{l_1}(\lambda) \times \nonumber \\
     && \mathcal{R}_\mathrm{d}(\lambda) \mathcal{T}_\mathrm{l_2}(\lambda) \mathcal{T}_\mathrm{f_1}(\lambda) \mathcal{T}_\mathrm{f_2}(\lambda) \mathrm{QE}(\lambda) \nonumber \\
    &=& \eta_A(\lambda,z)\eta_S(\lambda,z) \gamma_0(\lambda), 
\end{eqnarray}
where $\eta_A(\lambda,z)$ and $\eta_S(\lambda,z)$ are the absorption and scattering efficiencies in fiber (as described earlier), $\mathcal{T}(\lambda)$ and $\mathcal{R}(\lambda)$ are the transmittance and reflectance of an optic (w~=~window, l~=~lens, d~=~dichroic beamsplitter and f~=~spectral filter), and $\mathrm{QE}(\lambda)$ (cnt~photon$^{-1}$) is the quantum efficiency of the camera.

We describe the number of excitations per infinitesimal length $dz$ per pulse as
\begin{eqnarray}
\label{eq:Nexcit}
    &&N_C(z) = \frac{1}{2} \sigma_C n \times \\
    &&\int_{-1/2g}^{1/2g}\int_{-\infty}^{\infty}\int_{-\infty}^{\infty} \phi(x,y,z,t)^2 dxdydt, \nonumber
\end{eqnarray}
where $\sigma_C$ (1 GM = 10$^{-50}$ cm$^4$~s~photon$^{-1}$ fluorophore$^{-1}$) is the C2PA cross-section, $n$ (fluorophores~cm$^{-3}$) is the number density of the fluorophores and $\phi(x,y,z,t)$ (photons~cm$^{-2}$~s$^{-1}$) is the photon flux of the laser beam. The factor of $1/2$ carries units of excitations per photons absorbed. The temporal and transverse spatial profiles of the laser beam are approximated by Gaussian distributions. The transverse spatial profile of the light inside the fiber will differ from a Gaussian distribution if the light occupies the higher-order modes of the fiber, but in this calculation we make a few approximations based on the assumption that all the light is in the fundamental mode. The spatial integrals extend from negative infinity to positive infinity and are equivalent to integrals that extend over only the core of the fiber. Assuming the laser is always on, $\phi(x,y,z,t)$ takes the form
\begin{eqnarray}
\label{eq:flux}
&&\phi(x,y,z,t) = \phi_0(z)
\textrm{Exp}\left(-4\textrm{ln}2\left(\frac{x^2+y^2}{d_0 ^2}\right)\right) \nonumber \\
&&\times \sum_{i=-\infty}^{\infty} \textrm{Exp}\left(-4\textrm{ln}2\frac{(t+i/g)^2}{\tau(z)^2}\right) ,
\end{eqnarray}
where $\phi_0(z)$ (photons~cm$^{-2}$~s$^{-1}$) is the peak photon flux as a function of $z$, $\tau(z)$ (fs) is the FWHM pulse duration and $d_0$ (cm) is the FWHM beam width. We use COMSOL to solve for $d_0$ by first solving for the effective mode area $A_{\textrm{eff}}$ (cm$^2$) of the fundamental mode\cite{Agrawal2019} and solving
\begin{equation}
    d_0 = \sqrt{\frac{2\textrm{ln}2 A_{\textrm{eff}}}{\pi}}.
\end{equation}
We can define the average photon rate $Q(z)$ (photons~s$^{-1}$) in terms of the photon flux,
\begin{eqnarray}
\label{eq:avgPhotRate}
    Q(z) &=& g\int_{-1/2g}^{1/2g}\int_{-\infty}^{\infty}\int_{-\infty}^{\infty} \phi(x,y,z,t) dxdydt \nonumber \\
    &=& \frac{W(z)}{h\nu},
\end{eqnarray}
where $W(z)$ (W) is the average power of the beam as a function of $z$ and $h\nu$ (J) is the average energy of an incident photon. We can also write the power in a form to show its dependence on propagation losses in fiber as described in Eq.~\eqref{eq:avgPower}, which brings about its $z$-dependence.
\begin{table*}[!t]
%    \vspace*{0.2cm}
    \caption{Summary of variable parameters. Experiments 1 and 2 (Exp 1 and 2) use the same fiber (fiber 1) and laser excitation. Experiment 3 (Exp 3), which involves both laser and SPDC excitation, uses another fiber (fiber 2).}
    \label{Tab:varParam}
%    \vspace*{-0.2cm}
%    \centering
    \renewcommand{\tabcolsep}{14pt}
    \begin{tabular*}{\textwidth}{ llcccc }
    \midrule
    Parameter & \multicolumn{1}{c}{unit} & Exp 1 & Exp 2 & \multicolumn{2}{c}{Exp 3} \\ 
    & & & & Laser & SPDC \\ \midrule
    $F_C/W_0^2$ & cnt~s$^{-1}\,\mu$W$^2$& $3.14\times10^{3}$ & $3.19\times10^{4}$ & $3.62\times10^{5}$ & N/A \\
    $c$ & mM & $1.95\times10^{-2}$ & $1.70\times10^{-1}$ & \multicolumn{2}{c}{$2.30$} \\
    $l$ & cm & \multicolumn{2}{c}{37} & \multicolumn{2}{c}{36} \\
    $\mu(\lambda_e)$\textsuperscript{\emph{a}} & cm$^{-1}$ & \multicolumn{2}{c}{0.093} & \multicolumn{2}{c}{0.034} \\
    $D_0$ & fs$^2$ & 2000 & 0 & 0 & 2100 \\
    $\gamma_0(\lambda_f)$\textsuperscript{\emph{a}} & & \multicolumn{2}{c}{0.669} & \multicolumn{2}{c}{0.630} \\
    $\eta_T$ & & 0.40 & 0.43 & 0.47 & 0.43\\
    
    \midrule
    \end{tabular*}
    \\ \textsuperscript{\emph{a}} $\lambda_e=810$~nm, $\lambda_f=451$~nm 
\end{table*}

We can solve for $\phi_0(z)$ by inputting Eq.~\eqref{eq:flux} into Eq.~\eqref{eq:avgPhotRate} and integrating over $x$, $y$ and $t$,
\begin{equation}
\phi_0(z) = \left(\frac{4\textrm{ln}(2)}{\pi}\right)^{3/2}\frac{W(z)}{h\nu g d_0^2 \tau(z)}.
\end{equation}
The pulse duration has $z$ dependence because of dispersion and can be described by
\begin{equation}
\tau(z) = \sqrt{\tau_0^4+(4\textrm{ln}2)^2(D_0+\beta z)^2}/\tau_0 ,
\end{equation}
where $D_0$ (fs$^2$) is the GDD accumulated by the pulse before the fiber and $\beta$ (fs$^2$~cm$^{-1}$) is the total GVD accumulated by the light in the fundamental mode of the fiber. 

Now we can rewrite Eq.~\eqref{eq:FluRateFiber} using these equations as
\begin{eqnarray}
\label{eq:F_C}
&F_C = \sqrt{2} \left(\frac{\mathrm{ln}(2)}{\pi}\right)^{3/2} \frac{\sigma_C n W_0^2}{g (h \nu)^2 d_0^2} \times &\\
&\int_0^l \frac{\eta_A^2(\lambda_e,z)\eta_S^2(\lambda_e,z)}{\tau(z)} \int_{\lambda_1}^{\lambda_2} \gamma(z,\lambda) \kappa(\lambda) \Phi(\lambda) d\lambda dz .& \nonumber
\end{eqnarray}
where $\lambda_e = 810$~nm.

Then we can solve for the C2PA cross-section, 
\begin{eqnarray}
\label{eq:C2PAxSectionFiber}
&&\sigma_C = \frac{1}{\sqrt{2}}\left(\frac{\pi}{\mathrm{ln}(2)}\right)^{3/2} \frac{g (h \nu)^2 d_0^2}{n} \frac{F_C}{W_0^2} \nonumber\\
&& \times \left (\int_0^l \eta_A^2(\lambda_e,z)\eta_S^2(\lambda_e,z)/\tau(z) 
\right .  \\
&& \times \left . \int_{\lambda_1}^{\lambda_2} \gamma(z,\lambda) \kappa(\lambda) \Phi(\lambda) d\lambda dz \right)^{-1}. \nonumber
\end{eqnarray}

All the parameters in Eq.~\eqref{eq:C2PAxSectionFiber} are known through experiments or simulations. The parameters $d_0$ and $\beta$ are estimated using COMSOL Multiphysics Simulation software, $F_C/W_0^2$ (cnt~s$^{-1}$~$\mu$W$^{-2}$) is the fit to our experimental C2PEF data (and $W_0$ is estimated using measured $W(l)$), $n$ is measured using a spectrophotometer, $a_\mathrm{sol}(\lambda)$ is from literature\cite{Kedenburg2012}, $\mu(\lambda)$ is determined from scattering measurements, $\tau_0$ is measured using a SwampOptics Grenouille 8-50-USB, $D_0$ is measured, $k(\lambda)$ is calculated, $\Phi(\lambda)$ is known from published AF455 measurements\cite{Rogers2004,deReguardati2016} and $\gamma(z,\lambda)$ is calculated based on optics' specifications, our scattering measurements and the published extinction coefficient and spectra of AF455\cite{deReguardati2016}. Table~\ref{Tab:fixedParam} and Table~\ref{Tab:varParam} summarize parameter values used in this calculation and that described in the following section.

The uncertainty on derived C2PA cross-sections was determined by propagating the errors of the various parameters that go into Eq.~\eqref{eq:C2PAxSectionFiber}. We multiply this value by a coverage factor ($k=2$) to reach $\approx$~95\% confidence that the true value lies within the bounds set by the error bars. The uncertainty from each experiment is 34\%.

\begin{figure*}[!tbh]
\centering
\includegraphics[width=0.60\textwidth]{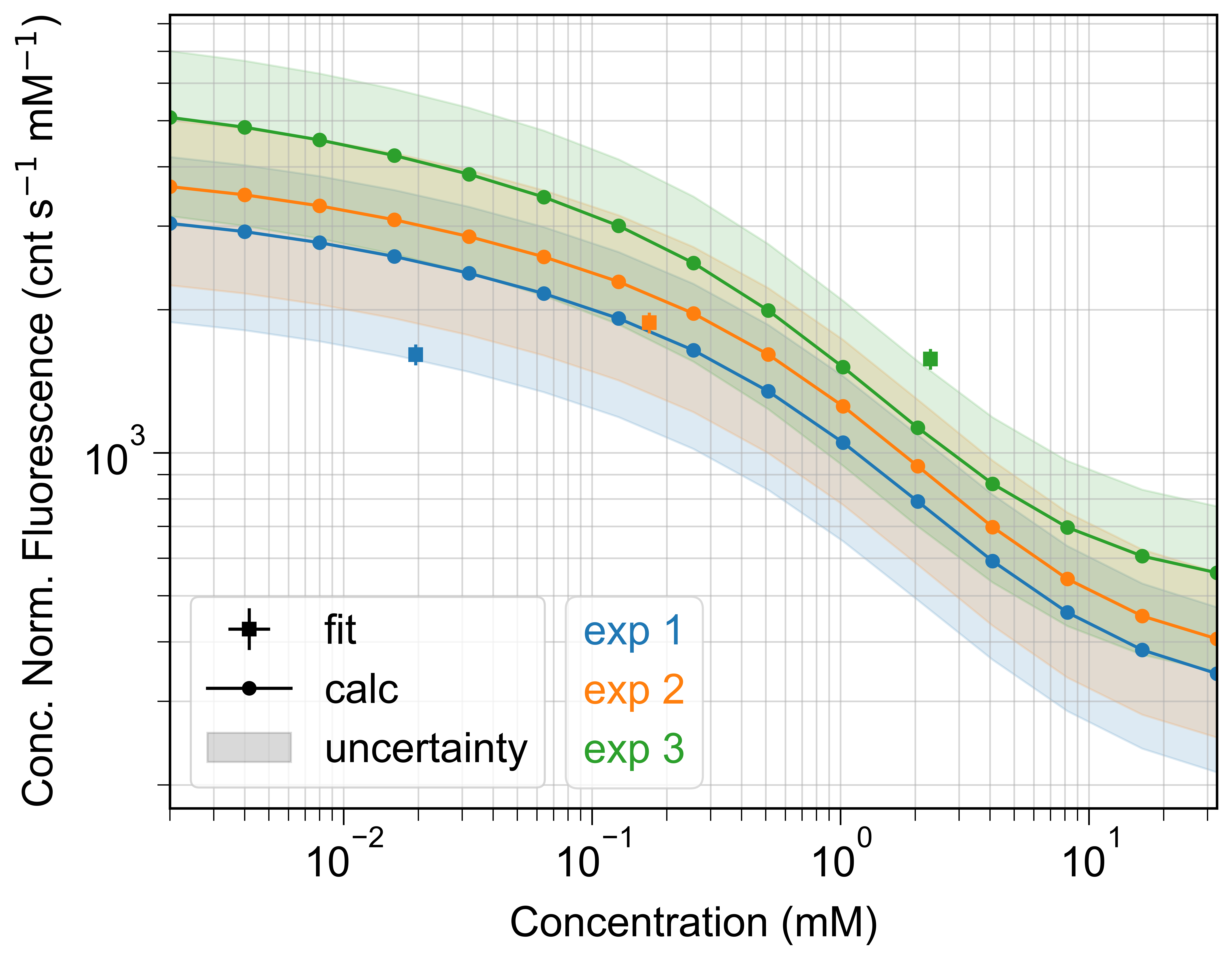}
\caption[simulation of concentration normalized fluorescence]{\label{fig:ConcNormFlu}Measured (fit) and calculated (calc) concentration (conc.) normalized (norm.) classical two-photon excited fluorescence at 100~nW excitation as a function of concentration. Experiments 1, 2 and 3 are shown in blue, orange and green respectively. Calculations (using Eq.~\eqref{eq:F_C}) were done for each experiment because of minor differences in the setup, which shifts the curve slightly. The calculations predict that as concentration increases there is a general decrease in the concentration normalized fluorescence, which is a result of reabsorption of fluorescence. A shaded region around each of the calculated results show the uncertainty in the calculation. For each experiment, the fit to the measured data was used to extract the concentration normalized fluorescence at 100~nW.}
\end{figure*}

In Figure~\ref{fig:ConcNormFlu}, we plot concentration normalized C2PEF ($\frac{F_C}{c}$) at $100$~nW from measurements (fit) and calculations (calc) for the experimental conditions of experiments 1 (blue), 2 (orange) and 3 (green). The calculations use Eq.~\eqref{eq:F_C} with the values of parameters input from Tables~\ref{Tab:fixedParam} and \ref{Tab:varParam}. The uncertainty on these values is propagated, and multiplied by a coverage factor ($k=2$) to give a total uncertainty of 38\% on the calculation (shown by shaded regions). After each experiment, small improvements to the setup were made that resulted in a shift of the curve upwards on the plot as experiments progressed from 1 to 3. The trend of each curve shows the same decrease of $\frac{F_C}{c}$ as a function of $c$. This general decrease is due to fluorescence reabsorption by the abundant molecules available at high concentrations. 
%At low concentrations this is irrelevant, but as the concentration is increased it slowly becomes more prevalent until it is so frequent that no fluorescence can make it through the fiber beyond a certain length. Then once the fluorescence at various lengths through the fiber is irrelevant, the primary source of fluorescence is right at the fiber entrance, which is also slowly attenuated.
The data points from measurements are extracted using quadratic fits (slope fixed to 2.00) to each experiments' dataset of fluorescence as a function of power. The fit is extrapolated to $100$~nW and that value is divided by the measured concentration of the experiment's sample. These data points are within the expected range based on the calculations and their respective uncertainties. 

\section*{Calculating an E2PA Cross-Section Upper Bound}
\label{Sec:FibE2PAXSection}
In this section, in a similar manner to the previous section, we discuss the assumptions and equations used to calculate an E2PEF signal. We arrive at an equation to derive an E2PA cross-section upper bound.

We model the E2PEF signal detected on the camera, $F_E$ (cnt~s$^{-1}$), as
\begin{equation}
\label{eq:EFluRateFiber}
    F_E = g \int_0^l N_E(z) \int_{\lambda_1}^{\lambda_2} \gamma(z,\lambda) \kappa(\lambda) \Phi(\lambda) d\lambda dz ,
\end{equation}
where $N_E(z)$ (excitations~cm$^{-1}$~pulse$^{-1}$) is the number of excitations per infinitesimal length of fiber $dz$ (cm) per pulse and the other parameters are as mentioned in the previous section. The parameter $N_E(z)$ is defined as
\begin{equation}
\label{eq:EntangNexcit}
    N_E(z) = \sigma_E(z)n\frac{Q_\mathrm{pairs}(z)}{g}
\end{equation}
where $\sigma_E(z)$ (cm$^2$~fluorophore$^{-1}$) is the E2PA cross-section, which has $z$ dependence due to group delay dispersion broadening the entanglement time and $Q_\mathrm{pairs}(z)$ (photon pairs~s$^{-1}$) is the SPDC photon pair rate in fiber. The other parameters are as mentioned in the previous section. 
%$\mathcal{T}$ is the free-space transmission efficiency of the SPDC from the center of the crystal to the fiber, and $\eta_T$ and $\eta_C$ are the fiber transmission and coupling efficiencies (see earlier definitions). 

In contrast to C2PA, E2PA requires the presence of spatially and temporally correlated photon pairs. Furthermore the excitation rate scales linearly, as opposed to quadratically, with photon rate in the low-gain regime. In this section, we use the scaling in the low-gain regime to make the most conservative estimate of the upper bound of the E2PA cross-section. In order to write an equation for $Q_\mathrm{pairs}(z)$, we will first write the SPDC single photon rate at the fiber, $Q(z)$ (photons~s$^{-1}$),
\begin{eqnarray}
    Q(z) &=& \eta_F\eta_C\eta_A(\lambda_e,z)\eta_S(\lambda_e,z)\frac{Q^{\mathrm{mm}}_\mathrm{xtal}}{M} \\
    &=& \eta_F\eta_C\eta_A(\lambda_e,z)\eta_S(\lambda_e,z)Q_\mathrm{xtal}, \nonumber
\end{eqnarray}
where $Q^{\mathrm{mm}}_\mathrm{xtal}$ (photons~s$^{-1}$) is the total (all modes) single photon rate at the crystal, $M$ is the number of spatial modes of the generated SPDC, $Q_\mathrm{xtal}$ (photons~s$^{-1}$) is the single photon rate at the crystal for the spatial mode coupled in fiber, and the other parameters are as mentioned in previous sections.

We can then write the SPDC photon pair rate at the fiber in terms of the single photon rate,
\begin{eqnarray}
Q_\mathrm{pairs}(z) &=& \eta_K\eta_A(\lambda_e,z)\eta_S(\lambda_e,z)\frac{Q(z)}{2} \\
&=& \eta_K\eta_F\eta_C\eta_A^2(\lambda_e,z)\eta_S^2(\lambda_e,z) \frac{Q_\mathrm{xtal}(z)}{2} \nonumber \\
&=& \eta^{'}_K\eta_F^2\eta_C^2\eta_A^2(\lambda_e,z)\eta_S^2(\lambda_e,z) \frac{Q_\mathrm{xtal}(z)}{2}, \nonumber
\end{eqnarray}
where $\eta_K^{'}$ is the effective Klyshko efficiency and $\eta_K$ is the Klyshko efficiency as defined in Eq.~\eqref{eq:Klyshko}. The third line of this equation illuminates a signature of E2PA---while E2PA scales linearly with photon rate, it scales quadratically with single photon loss. More details on this signature can be found in Ref.~\citenum{Parzuchowski}. 

Here we will use the probabilistic model (see for example Ref.~\citenum{Parzuchowski}) to describe $\sigma_E(z)$ as
\begin{equation}
\sigma_E(z) \approx \frac{\sigma_C}{T_e(z)A_e},
\end{equation}
where $T_e(z)$ (fs) is the entanglement time which increases in the presence of group velocity dispersion and thus depends on $z$ and $A_e$ (cm$^2$) is the entanglement area. 

%In accordance with $T_e$ depending inversely on the bandwidth of the SPDC\cite{Raymer2021}, we approximate the $z$-dependence of $T_e$ using
%\begin{equation}
%    T_e(z) \approx \sqrt{T_{e,0}^4+(4\mathrm{ln}2)^2(\beta z+D_0)^2}/T_{e,0}
%\end{equation}
%where $T_{e,0}$ (fs) is the entanglement time of the SPDC at the center of the crystal and is estimated to be 17~fs based on the discrete Fourier transform of the measured SPDC spectrum shown in Ref.~\citenum{Parzuchowski}. The parameters $D_0$ and $\beta$ are described in Section~\ref{Sec:FibC2PAXSection}. 
We simulate the entanglement time using a discrete Fourier transform of the measured SPDC joint spectrum shown in Ref.~\citenum{Parzuchowski}. To do this, we estimate the joint spectral amplitude as
\begin{eqnarray}
    f(\omega_{\mathrm{S}},\omega_{\mathrm{I}}, z)= \sqrt{F(\omega_{\mathrm{S}},\omega_{\mathrm{I}})} \\
    \times \mathrm{Exp}\left(i(D_0+\beta z)(\omega_{\mathrm{S}}-\omega_{\mathrm{P}}/2)^2/2\right) \nonumber \\
    \times \mathrm{Exp}\left(i(D_0+\beta z)(\omega_{\mathrm{I}}-\omega_{\mathrm{P}}/2)^2/2\right), \nonumber
\end{eqnarray}
where $F(\omega_{\mathrm{S}},\omega_{\mathrm{I}})$ is the measured joint spectral intensity, and $\omega_{\mathrm{S}}$, $\omega_{\mathrm{I}}$ and $\omega_{\mathrm{P}}$ are the signal, idler and pump frequencies and the parameters $D_0$ and $\beta$ are described in the previous section. Discrete Fourier transforms are performed at discrete steps along the fiber of 1~cm and the entanglement time is calculated as the FWHM of the projection of the joint temporal intensity along the ($t_{\mathrm{S}}-t_{\mathrm{I}}$) axis. Here $t_{\mathrm{S}}$ and $t_{\mathrm{I}}$ are the time of arrival of signal and idler respectively. The FWHM is calculated from standard deviation due to the spectrum's non-Gaussian nature. We fit the $z$-dependence of the resulting $T_e$ values using
\begin{equation}
\label{eq:Te}
    T_e(z) = 2\sqrt{2\mathrm{ln}2}\sqrt{T_{e,0}^4+S_0(\beta z+D_0)^2}/T_{e,0},
\end{equation}
where $T_{e,0}$ (fs) and $S_0$ are fitting parameters and the preceding numerical factors convert standard deviation to FWHM. Physically $T_{e,0}$ should correspond to the entanglement time of the SPDC at the center of the crystal, which is 17~fs, however for our conditions that would render the fit far from ideal. We believe the reason for this is related to the very non-Gaussian form of the spectrum. The parameter $S_0$ is a numerical factor that is also related to the shape of the spectrum. 

In order to estimate a single value for the E2PA cross-section, we will use
\begin{equation}
    \sigma_E(0) = \frac{\sigma_C}{T_e(0)A_e}.
\end{equation}
We can then rewrite Eq.~\eqref{eq:EntangNexcit} as
\begin{equation}
    N_E(z) = \sigma_E(0)n\frac{T_e(0)}{T_e(z)}\frac{Q_\mathrm{pairs}(z)}{g}.
\end{equation}
Then we can rewrite Eq.~\eqref{eq:EFluRateFiber} as
\begin{eqnarray}
    &&F_E = \sigma_E(0) n \int_0^l \frac{T_e(0)}{T_e(z)} Q_\mathrm{pairs}(z) \nonumber \\
    &&\times \int_{\lambda_1}^{\lambda_2} \gamma(z,\lambda) \kappa(\lambda) \Phi(\lambda) d\lambda dz.
\end{eqnarray}
We can solve for $\sigma_E(0)$,
\begin{eqnarray}
&&\sigma_E(0)= F_E \times \left ( n \int_0^l \frac{T_e(0)}{T_e(z)} Q_\mathrm{pairs}(z) \right . \nonumber \\
&&\left . \times \int_{\lambda_1}^{\lambda_2} \gamma(z,\lambda) \kappa(\lambda) \Phi(\lambda) d\lambda dz \right )^{-1}.
\end{eqnarray}
Then for a null measurement, we replace $F_E$ by the measurable fluorescence lower bound $F^{\mathrm{LB}}$ (cnt~s$^{-1}$) and $\sigma_E(0)$ becomes $\sigma_E^{\mathrm{UB}}$,
\begin{eqnarray}
\label{eq:E2PAxSectionUB}
&&\sigma_E^{\mathrm{UB}}= F^{\mathrm{LB}} \times \left ( n \int_0^l \frac{T_e(0)}{T_e(z)} Q_\mathrm{pairs}(z) \right . \nonumber \\
&&\left .\times \int_{\lambda_1}^{\lambda_2} \gamma(z,\lambda) \kappa(\lambda) \Phi(\lambda) d\lambda dz  \right )^{-1}.
\end{eqnarray}
All the parameters in Eq.~\eqref{eq:E2PAxSectionUB} are known through experiments or simulations. The parameter $F^{\mathrm{LB}}$ is estimated to be $1.0$~cnt~s$^{-1}$ based on the signal levels that could be distinguished in these measurements, $T_{e,0}$ and $S_0$ are estimated using the discrete Fourier transform of the measured joint spectrum, $D_0$ is estimated, $Q_\mathrm{pairs}(z)$ is calculated using the measured value of $Q_\mathrm{out}$ and measured and estimated values of efficiencies, and all other parameters are found as stated in the previous section. The values of these parameters are listed in Tables~\ref{Tab:fixedParam} and \ref{Tab:varParam}.

The uncertainty on the derived E2PA cross-section upper bound was determined by propagating the errors of the various parameters that go into Eq.~\eqref{eq:E2PAxSectionUB} and multiplying this value by a coverage factor ($k=2$), which gives 40\%.


\bibliography{fiberPaper}

\end{document}